\documentclass[%
reprint,
superscriptaddress,
amsmath,amssymb, 
aps,
article,pre,
]{revtex4-2}

\usepackage[normalem]{ulem}
\usepackage{cancel}
\usepackage{graphicx}
\usepackage{dcolumn}
\usepackage{bm}

\usepackage[T1]{fontenc}
\usepackage{mathptmx}
\usepackage{etoolbox}

\usepackage{enumerate}

\usepackage{cancel}
\usepackage{color}

\usepackage[colorlinks=true,citecolor=black,linkcolor=black,urlcolor=blue]{hyperref}
\usepackage{xr-hyper}
\usepackage{physics}

\newcolumntype{P}[1]{>{\centering\arraybackslash}p{#1}}

\newcommand{\kT}{k_{\rm B}T}
\newcommand{\tR}{\tau_{\rm r}}
\newcommand{\tDiff}{\tau_{\rm diff}}
\newcommand{\tSW}{\tau_{\rm res}}

\DeclareMathAlphabet{\vmath}{OT1}{pzc}{m}{it}
\DeclareSymbolFont{pazo-cal}{OMS}{fplm}{m}{n}
\DeclareSymbolFontAlphabet{\mathcal}{pazo-cal}
\begin{document}

\title{Information-driven stepping in dimeric transport motors}

\author{Antonio Patrón Castro}%
 \email{apatronc@sfu.ca}
 \affiliation{Department of Physics, Simon Fraser University, Burnaby, British Columbia V5A 1S6, Canada}
  \author{David A. Sivak}%
 \email{dsivak@sfu.ca}
 \affiliation{Department of Physics, Simon Fraser University, Burnaby, British Columbia V5A 1S6, Canada}

\begin{abstract}
Dimeric transport motors are nanoscale protein complexes that move along cytoskeletal filaments. Here we introduce a theoretical model for their stepping dynamics, in which the two motor heads undergo Brownian motion with mobilities periodically switching in a position-dependent manner so that only one head moves at a time. Through consumption of chemical free energy, this mechanism produces directed motion. We characterize at steady state the motor's mean velocity and its energy and information flows. The motor operates as a pure information engine, where only information is transduced between its components, without energy exchange. For localized switching, the model yields a thermodynamically consistent expression for the mean motor velocity that reproduces experimentally observed behavior, and captures the stall force characteristic of tightly coupled motors. Finally, coarse-graining the model to a single mechanical degree of freedom produces a second-order non-Markovian dynamics, from which we compute the four distinct dwell-time distributions that can be directly observed in single-molecule experiments. Our findings highlight how information transduction, via implicit operation as a Maxwell demon, may underlie the remarkable performance of these molecular motors.
\end{abstract}

\pacs{Valid PACS appear here}
\maketitle

\section{\label{sec:intro}Introduction}
Biology serves as a fertile ground for exploring nonequilibrium phenomena. Biological systems continually fight the second law of thermodynamics to maintain complex, ordered structures and to carry out the myriad tasks required to sustain life. Within living organisms, cells are kept out of equilibrium by molecular machines---nanoscale protein structures that convert between different forms of free energy~\cite{brown_theory_2019}. Drawing on nonequilibrium flows that ultimately originate from solar photons~\cite{vinyard_photosystem_2013}, these molecular machines perform diverse functions and uphold the low-entropy organization of cells~\cite{alberts_cell_1998}. Unlike conventional macroscopic engines, however, molecular machines face unique challenges: their dynamics are highly overdamped, dominated by frictional drag rather than inertia~\cite{brown_theory_2019,purcell_life_2014,bustamante_physics_2001}; they are strongly influenced by stochastic environmental fluctuations; and their protein architectures incorporate flexible components joined by loose, compliant connections~\cite{brown_toward_2017}. Despite these challenges, molecular machines exhibit remarkable performance---often rivaling or surpassing that of their macroscopic counterparts~\cite{marden_molecules_2002}---frequently achieving both high energetic efficiency and rapid operation~\cite{leighton_flow_2025,ito_kinetic_2007,berg_rotary_2003}.

\begin{figure}
  \centering
  \includegraphics[width=0.9\columnwidth,height=1.5\columnwidth]{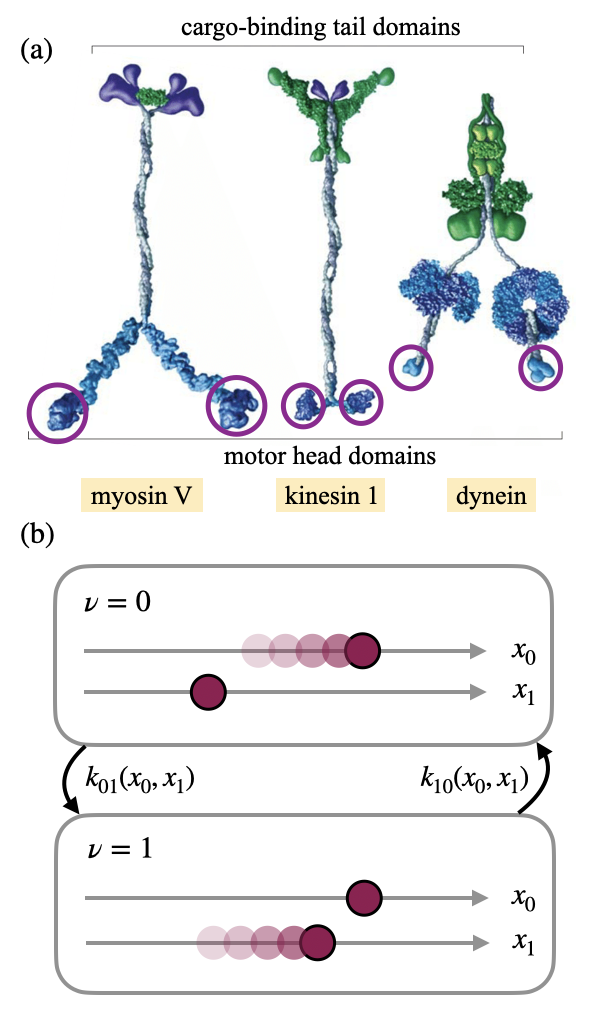}
  \caption{(a) Key classes of transport motors, adapted from~\cite{vale_molecular_2003}.Purple circles indicate the motor domains (heads), whose motion is modeled via the coupled Langevin equations~\eqref{eq:langevin}. (b) Switching-mobility model for the heads of a dimeric transport motor. The motor is represented by two overdamped Brownian particles in one dimension, with positions $x_0$ and $x_1$, whose mobilities switch according to a discrete internal chemical state $\nu$. For $\nu = 0$, $x_0$ evolves while $x_1$ is fixed; for $\nu = 1$, $x_1$ evolves while $x_0$ is fixed.}
  \label{fig:sketch-model}
\end{figure}

A prominent class of molecular machines is the transport motors~\cite{schliwa_molecular_2003,kolomeisky_molecular_2007}. These nanoscale engines, which include the kinesin~\cite{block_kinesin_2007}, myosin~\cite{spudich_myosin_2001}, and dynein~\cite{bhabha_how_2016} superfamilies, convert chemical free energy---most commonly from the hydrolysis of adenosine triphosphate (ATP) to adenosine diphosphate (ADP) and inorganic phosphate---into mechanical work that drives the transport of cargoes (e.g., vesicles, organelles, mRNA, and proteins) along cytoskeletal structures~\cite{hackney_kinetic_1996,woehlke_microtubule_1997,howard_mechanics_2002,spudich_myosin_2010}. Beyond carrying cargo, these motors are essential for a variety of intracellular processes, including muscle contraction~\cite{rastogi_maximum_2016,spudich_myosin_2010} and chromosome segregation in mitosis~\cite{howard_mechanics_2002}. Of special interest is the subclass of dimeric transport motors, composed of two motor domains (or heads) that operate in a coordinated fashion to generate directed motion along filaments (Fig.~\ref{fig:sketch-model}a). Unlike monomeric motors, whose displacement arises from cycles of binding to and complete detachment from the track~\cite{laakso_myosin_2008,schimert_intracellular_2019}, dimeric motors step by advancing one head at a time while the other head remains firmly attached, so that the motor typically does not completely detach from the track. This alternating-head mechanism thus endows these motors with high processivity, enabling them to traverse long distances along the cytoskeleton~\cite{howard_mechanics_2002,leibler_porters_1993}.

There has been extensive research on modeling molecular transport motors within the framework of nonequilibrium statistical mechanics. The models employed can be classified into two broad types. On the one hand, in continuous flashing-ratchet models~\cite{peskin_coordinated_1995,julier_modeling_1997,keller_mechanochemistry_2000,bustamante_physics_2001,reimann_brownian_2002,brown_pulling_2019,xing_making_2005,kanada_theoretical_2003}, a few collective coordinates describe the motor's major conformational movements. Motion is diffusive, with forces derived from asymmetric free-energy landscapes governing each coordinate, and with jumps between these landscapes corresponding to chemical transitions~\cite{xing_making_2005}. On the other hand, purely discrete models based on master equations~\cite{valleriani_dwell_2008,leibler_porters_1993,kolomeisky_simplified_1998,fisher_molecular_1999,kolomeisky_periodic_2000,kolomeisky_extended_2000,kolomeisky_exact_2001,fisher_simple_2001,kolomeisky_simple_2003,stukalin_mechanochemical_2005,stukalin_coupling_2005,fisher_kinesin_2005,kim_vectorial_2005,qian_cycle_2005,stukalin_transport_2006} have been widely employed for motors such as kinesin-1 and myosin-V, which advance in steps of fixed length along their tracks. In this approach, the stepping mechanics is coarse-grained to a discrete level, and each transition between discrete states represents a mechanochemical process, such as ATP binding, ADP release, or forward or backward stepping. 

Stochastic thermodynamics provides a framework to study how transport motors functionally transduce different forms of free energy. Beyond the exchange of internal energy between a motor's multiple components, recent developments in the field have highlighted the role of information as a thermodynamic resource~\cite{landauer_information_1991,parrondo_thermodynamics_2015}. This gives rise to the concept of information engines~\cite{szilard_on_1964}, where directed motion and work extraction can be achieved through the transduction of information alone, without energetic exchange, analogous to the operation of Maxwell's  demon. In such a system, creation of correlations among different degrees of freedom effectively replaces energy transmission between them. Experimental realizations of information engines have largely relied on non-autonomous systems~\cite{toyabe_experimental_2010,camati_experimental_2016,koski_on-chip_2015,cottet_observing_2017,masuyama_information_2018,koski_experimental_2014,chida_power_2017,admon_experimental_2018,goerlich_experimental_2025,baldovin_optimal_2025,patron_castro_harnessing_2026}, in which an external time-dependent controller uses feedback to capitalize on correlations. Understanding whether autonomous molecular systems can exploit similar information-transduction mechanisms, and how these mechanisms compare with energetic driving, remains an open and compelling problem~\cite{tsuruyama_rna_2023,leighton_inferring_2023,leighton_flow_2025}.

In this work, we introduce a stochastic model for the stepping dynamics of a generic dimeric transport motor, solely driven by information transduction without energy exchange. We model the motor by coupled overdamped Brownian motion of the two individual heads, with mobilities alternating periodically so that at any given time only one head is mobile. The dynamics takes place under a constant opposing force (accounting for the cargo load) and a pairwise interaction potential between the heads. In contrast to traditional flashing-ratchet models, directionality does not originate from biased free-energy landscapes; instead, it arises from asymmetric head-to-track interactions modeled via position-dependent chemical 
switching rates. The nonequilibrium driving force is provided by the chemical free energy consumed per step, which ultimately dictates the magnitude of this directional bias. In fact, the total mechanical energy landscape (pairwise potential plus effects of the cargo force) does not favor forward motion, and it does not change upon the chemical switches. Consequently, each head must first reach specific target positions through diffusional search before being locked in by a chemical transition enabling mobility exchange. Thermodynamically, the motor operates as an information engine, in which the chemical degree of freedom implicitly acts as a Maxwell demon controlling the moving heads.

We analyze the long-time steady-state dynamics and characterize mechanical observables such as the motor’s mean velocity, a single motor head's mean step distance, and the mean step duration, together with the motor's thermodynamic flows. For localized switching, all relevant quantities admit analytical expressions in terms of physically meaningful quantities, including the steady-state rates of forward and backward switches, the step length, and the characteristic switching times. Finally, also for localized switching, we coarse-grain the model to obtain an effective discrete-state description, and compute the dwell-time distributions associated with forward and backward stepping, which are directly accessible in single-molecule experiments. Overall, our findings show that information transduction alone provides a framework to understand how the dynamics of dimeric motors are produced by chemical driving.

\section{\label{sec:model}Coupled Brownian particles with switching mobilities}
We represent the motion of the two motor heads of a dimeric transport motor by two overdamped Brownian particles (Fig.~\ref{fig:sketch-model}b shows a schematic) with one-dimensional mechanical coordinates $X=(x_0,x_1)$ (i.e., along the cytoskeletal filament), whose dynamics obeys the coupled Langevin equations
\begin{subequations}\label{eq:langevin}
\begin{align}
    \dot{x}_0 &= \delta_{\nu, 0} \left[-\beta D(\partial_{x_0} V + f) + \sqrt{2D}\ \xi_0(t)\right]    
    \\
    \dot{x}_1 &= \delta_{\nu, 1} \left[-\beta D(\partial_{x_1} V + f) + \sqrt{2D}\ \xi_1(t)\right]\ ,
\end{align}
\end{subequations}
for diffusion coefficient $D$ and inverse temperature $\beta \equiv (k_{\rm B}T)^{-1}$, where $k_{\rm B}$ is Boltzmann's constant and $T$ is the temperature of the thermal bath. The attractive pairwise potential $V = V(x_0-x_1)$ due to the linker between the heads depends solely and symmetrically on their relative displacement: $V(x_0-x_1) = V(x_1-x_0)$. The force $f \geq 0$ quantifies the opposing load exerted by the cargo on each head.

The terms $\xi_0(t)$ and $\xi_1(t)$ are independent Gaussian white-noise fluctuations with zero mean and correlations
\begin{equation}
\left<\xi_i(t)\, \xi_j(t')\right> = \delta_{ij}\, \delta (t-t')\ , \quad i,j=\{0, 1\} \ ,
\end{equation}
where $\left< \cdot \right>$ denotes the average over noise realizations. Finally, $\nu = \{0,1\}$ is a two-state chemical variable enforcing that only one head moves at a time. For $\nu = 0$, $x_0$ evolves while $x_1$ remains fixed, whereas for $\nu = 1$, the roles are reversed. This captures the coordination of the motor heads: one head remains attached to the filament while the other searches for a new binding position. This model draws inspiration from the model presented in~\cite{ehrich_energetic_2023}, which was used to study the information costs in the joint dynamics of a system and its feedback controller.

We introduce the mechanical scales defining the natural units of the problem, based on the long-time equilibrium behavior of one head when the other head is held fixed. The mechanical length scale is the equilibrium standard deviation $\sigma_{\text{eq}}$ of the head's position, while the mechanical time scale is the relaxation time $\tR$ in the potential. The characteristic energy scale is set by the thermal energy $\kT$. For a harmonic linker potential $V(x_0-x_1) = \tfrac{1}{2} \kappa (x_0-x_1)^2$, the characteristic length is $\sigma_{\text{eq}} = (\beta \kappa)^{-1/2}$, and $\tR = (\beta D \kappa)^{-1}$.  Thus the characteristic force scale is the drag force $f^* \equiv \sigma_{\text{eq}}(\beta D\tR)^{-1}$ required to move one motor head by one equilibrium standard deviation over one relaxation time. 

The mobility-switching mechanism captures the high processivity of dimeric transport motors: for the motor to step forward, one head must remain firmly attached to the filament while the other searches for a new binding site. The switching dynamics of the mobility variable $\nu$ represents the underlying chemical processes that regulate head attachment and detachment to the filament and, in combination with spatial asymmetry, enable biased motion even against the cargo force. The model does not explicitly include interactions between the motor heads and the filament; rather, the filament dictates the switching probability as a function of the motor-head positions and constrains one head at a time to a fixed position along it. The pairwise potential models the elastic coupling between the motor heads due to the linker, limiting their relative separation. Since the total energy landscape (involving the linker potential and the effects of the cargo) makes reaching the forward binding site energetically unfavorable, the moving motor head relies on diffusional search to reach the forward region where mobility can switch. 

At the ensemble level, these dynamics satisfy the coupled Fokker-Planck equations with source terms,
\begin{subequations}\label{eq:fokker-planck}
\begin{align}
\partial_t P_t(0,x_0,x_1) &= \mathbb{L}_f^{(x_0)}[P_t(0,x_0,x_1)] - k_{01}(x_0,x_1)\, P_t(0,x_0,x_1) \nonumber
\\
& \quad + k_{10}(x_0,x_1)\, P_t(1,x_0,x_1)\ 
\\
\partial_t P_t(1,x_0,x_1) &= \mathbb{L}_f^{(x_1)}[P_t(1,x_0,x_1)] + k_{01}(x_0,x_1)\, P_t(0,x_0,x_1) \nonumber
\\
& \quad - k_{10}(x_0,x_1)\, P_t(1,x_0,x_1)\ 
\end{align}
\end{subequations}
for the Fokker-Planck operator at fixed $f$, 
\begin{equation}\label{eq:FP-operator}
    \mathbb{L}_f^{(x)}[\phi] \equiv  \partial_x \left[\beta D(\partial_xV + f)\phi + D\, \partial_x \phi \right] \ .
\end{equation}
The joint probability $P_t(\nu,x_0,x_1)$ satisfies the normalization condition
\begin{equation}
    \sum_{\nu=0}^1\int_{-\infty}^{+\infty}{\rm d}x_0 \int_{-\infty}^{+\infty}{\rm d}x_1  \ P_t(\nu,x_0,x_1) = 1 \quad \forall t \ .
\end{equation}
The source terms in Eqs.~\eqref{eq:fokker-planck} involve the respective transition rates $k_{01}(x_0,x_1)$ and $k_{10}(x_0,x_1)$ for the mobility switches $0 \to 1$ and $1 \to 0$. To ensure thermodynamic consistency, these rates satisfy the local detailed-balance condition
\begin{equation}\label{eq:detailed-balance}
    \log \frac{k_{{01}}(x_0,x_1)}{k_{{10}}(x_0,x_1)} = \beta\Delta\mu\,h(x_0-x_1) \ ,
\end{equation}
for chemical free energy $\Delta\mu$, which constitutes a coarse-grained chemical driving associated with the relative concentrations of ATP, ADP, and phosphate. The function \( h(x_0-x_1) \) is a continuous odd function of the relative displacement \( x_0-x_1 \) that rapidly saturates to \( \pm 1 \) with increasing $|x_0-x_1|$. The odd character of this function favors forward motion: in state \(0\), where only \(x_0\) moves, the transition \(0 \to 1\) is more likely when \(x_0 > x_1\) than when \(x_0 < x_1\). Conversely, in state \(1\), where only \(x_1\) moves, the transition \(1 \to 0\) is more likely when \(x_1 > x_0\) than when \(x_1 < x_0\). The overall effect is to generate positive net motion.

We further assume that the two motor heads are indistinguishable (forming a homodimer), such that $k_{{01}}(x_0,x_1)=k_{{10}}(x_1,x_0)$, and that both rates depend exclusively on the relative displacement $x_0-x_1$. Following these assumptions, the transition rates take the form
\begin{subequations}\label{eq:transition-rates}
\begin{align}
    k_{{01}}(x_0,x_1) &\equiv k(x_0-x_1) 
    \\
    &= \Gamma(x_0-x_1)\, \exp\{\tfrac{1}{2}\beta\Delta\mu\,h(x_0-x_1)\} \\ 
    k_{{10}}(x_0,x_1) &= k(x_1-x_0)
    \\
    &= \Gamma(x_1-x_0)\, \exp\{\tfrac{1}{2}\beta\Delta\mu\,h(x_1-x_0)\} \ .
\end{align}
\end{subequations}
Here $\Gamma(x_0-x_1)$ is the switching rate with no chemical driving, an even function that quantifies the switching propensity as a function of head-head separation $x_0-x_1$. For a homodimer, the system of Fokker--Planck equations~\eqref{eq:fokker-planck} is invariant under the exchanges $x_0 \leftrightarrow x_1$ and $0 \leftrightarrow 1$, and therefore admits solutions satisfying $P_t(0,x_0,x_1)=P_t(1,x_1,x_0)$. 

Although in a homodimer the motor heads are indistinguishable, their interaction with the filament---represented by the switching dynamics of the $\nu$ variable---depends asymmetrically on the mobile head's position relative to the fixed head, with the chemical driving $\beta \Delta \mu$ quantifying the magnitude of the directional bias. In contrast with other flashing-ratchet models, where biased motion typically results from a combination of asymmetric free-energy landscapes and discrete-state transitions, our model's driving mechanism is asymmetric switches depending on the relative position of the heads. Since the combined effect of the linker potential and the opposing cargo force does not lead to average progress towards the forward binding site, diffusive exploration (i.e., random fluctuations) is necessary to reach the regions where switches occur.

\subsection{Relative displacement and motor position}
Our aim is to characterize the long-time steady-state behavior, corresponding to the regime in which molecular motors typically operate. Because the coordinates $(x_0,x_1)$ have a net average drift, the Fokker--Planck equations~\eqref{eq:fokker-planck} do not admit straightforward steady-state solutions; however, a suitable change of variables permits analysis of the long-time dynamics.

Motivated by the structure of the spatial dynamics~\eqref{eq:fokker-planck} and the switching rates~\eqref{eq:transition-rates}, we change variables to the relative displacement $r \equiv x_0-x_1$ and the center of mass (or motor position) $R \equiv \tfrac{1}{2}(x_0 + x_1)$. The Fokker-Planck equations~\eqref{eq:fokker-planck} become
\begin{subequations}
\begin{align}
\partial_t P_t(0,r,R) &= \mathbb{L}_{f}^{(r)}[P_t(0,r,R)] + D\ \partial_{rR}^2 P_t(0,r,R) \nonumber
\\
& \quad + \tfrac{1}{2}\partial_R \big[\beta D({\rm d}_r V + f)P_t(0,r,R) + \tfrac{1}{2}D \ \partial_R P_t(0,r,R)\big] \nonumber
\\
& \quad - k(r)P_t(0,r,R) + k(-r)P_t(1,r,R) 
\\
\partial_t P_t(1,r,R) &= \mathbb{L}_{-f}^{(r)}[P_t(1,r,R)] - D \ \partial_{rR}^2 P_t(1,r,R) \nonumber
\\
& \quad + \tfrac{1}{2}\partial_R \big[\beta D(f - {\rm d}_r V)P_t(1,r,R) + \tfrac{1}{2}D \ \partial_R P_t(1,r,R)\big] \nonumber
\\
& \quad + k(r)P_t(0,r,R) - k(-r)P_t(1,r,R) \ ,
\end{align}
\end{subequations}
for the probability $P_t(\nu,r,R)$ in the new variables $(r,R)$, satisfying the normalization
\begin{equation}\label{eq:normalisation}
    \sum_{\nu = 0}^1\int_{-\infty}^{+\infty}{\rm d}r 
     \int_{-\infty}^{+\infty}{\rm d}R \ P_t(\nu,r,R)=1 \quad \forall t \ .
\end{equation}

Integrating over the motor position $R$ yields the probability $P_t(\nu,r)$ for the relative coordinate $r$ and chemical state $\nu$. Such probability satisfies the reduced system of Fokker-Planck equations
\begin{subequations}\label{eq:PDE-marginal}
\begin{align}
\partial_t P_t(0,r) &= \mathbb{L}_{f}^{(r)}[P_t(0,r)] - k(r)P_t(0,r) 
+ k(-r)P_t(1,r) \label{eq:marginal-pdes-1}
\\
\partial_t P_t(1,r) &= \mathbb{L}_{-f}^{(r)}[P_t(1,r)] + k(r)P_t(0,r) 
- k(-r)P_t(1,r) \ ,
\end{align}
\end{subequations}
which admit steady-state solutions satisfying the symmetry $P_{\infty}(0,r)=P_{\infty}(1,-r)$, reflecting the invariance under exchange of the $x_0$ and $x_1$ coordinates and flipping the value of $\nu$. In SM I, we explore the consequences of this symmetry, which are relevant for the subsequent steady-state analysis. In the following, we will also work with the steady-state relative-displacement distribution conditional on the chemical state, 
\begin{subequations}
\begin{align} 
    P_{\infty}(r|\nu)
    &\equiv
    \frac{P_{\infty}(\nu,r)}
    {P_{\infty}(\nu)}
    \\
    &= 2 P_{\infty}(\nu,r) \, .
\end{align}
\end{subequations}
The factor of $2$ follows from the homodimer symmetry. This distribution characterizes the steady-state dynamics and provides the basis for computing the mean velocity and thermodynamic flows.

\subsection{Mean motor velocity}
While the distribution for relative coordinate $r$ eventually reaches steady state, the motor position $R$ drifts persistently and therefore has no steady-state solution. Its rate of change is
\begin{subequations}\label{eq:mean-vel-1}
\begin{align}
    {\rm d}_t\left<R\right>_t &= \sum_{\nu = 0}^1\int_{-\infty}^{+\infty}{\rm d}r\int_{-\infty}^{+\infty}{\rm d}R \ R \ \partial_t P_t(\nu,r,R) \label{eq:mean-vel-1-1}
    \\
    &= \frac{1}{4}\beta D\left[\left<{\rm d}_r V\right>^{(1)}_t - \left<{\rm d}_rV\right>^{(0)}_t - 2f\right] \ . \label{eq:velocity-ss}
\end{align}
\end{subequations}
The second line involves integration by parts. The quantities $\left<{\rm d}_rV\right>_t^{(0)}$ and $\left<{\rm d}_rV\right>_t^{(1)}$ are the respective averages of ${\rm d}_rV$ over $P_t(r|0)$ and $P_t(r|1)$. Equation~\eqref{eq:velocity-ss} is alternatively obtained from the Langevin equations~\eqref{eq:langevin} by taking into account that ${\rm d}_t\left<R\right>_t = \tfrac{1}{2}({\rm d}_t\left<x_0\right>_t + {\rm d}_t\left<x_1\right>_t)$. Averaging Eq.~\eqref{eq:velocity-ss} over $r$ yields the steady-state mean motor velocity, one of the key performance metrics.

A more physically intuitive expression for the mean motor velocity comes from the trajectory-level velocity of a detached motor head---for instance, the $x_0$ head during state $0$. Each trajectory can be viewed as a series of independent ratchet events, in which the $x_0$ head moves a displacement $\Delta x^{(0)}_{n}$ for a time $\Delta t_n^{(0)}$ and then remains fixed for a time ${\Delta t}_n^{(1)}$. For an infinitely long trajectory with $N_{\text{sw}} \to \infty$ switches, the velocity is
\begin{subequations}
\begin{align}
    \mathcal{V} &\equiv \lim_{N_{\text{sw}}\to \infty} \frac{N_{\text{sw}}^{-1} \, \sum_{n=0}^{N_{\text{sw}}}\Delta x^{(0)}_{n}}{N_{\text{sw}}^{-1} \, \sum_{n=0}^{N_{\text{sw}}}\left(\Delta t_n^{(0)} + \Delta t_n^{(1)}\right)}
    \\
    &=  \frac{\overline{\Delta x^{(0)}}}{\overline{\Delta t^{(0)}} + \overline{\Delta t^{(1)}}} \ ,
\end{align}
\end{subequations}
for the trajectory-averaged displacement $\overline{\Delta x^{(0)}}$ of the moving head ($x_0$), the trajectory-averaged durations $\overline{\Delta t^{(0)}}$ that $x_0$ remains mobile and $\overline{\Delta t^{(1)}}$ that it remains fixed. Due to homodimer symmetry, $\overline{\Delta x^{(0)}} = \overline{\Delta x^{(1)}} \equiv \overline{\Delta r}$ and $\overline{\Delta t^{(0)}} = \overline{\Delta t^{(1)}} \equiv \overline{\Delta t}$. The average motor displacement is $\overline{\Delta R} = \overline{\Delta r}/2$, thus yielding
\begin{equation}\label{eq:mean-vel-phys}
    \mathcal{V} = \frac{\overline{\Delta R}}{\overline{\Delta t}} \ .
\end{equation}
(In SM II, we show the equivalence of the two expressions \eqref{eq:velocity-ss} and \eqref{eq:mean-vel-phys} for the mean motor velocity.)

To obtain the mean motor displacement $\overline{\Delta R}$, we introduce 
\begin{equation}\label{eq:distribution-r-displ}
    {\rm P}_{\text{sw}}(r|\nu) \equiv \frac{k(r)P_{\infty}(r|\nu)}{\int_{-\infty}^{\infty}{\rm d}r\  k(r)P_{\infty}(r|\nu)} \ ,
\end{equation}
the steady-state probability that the next switch occurs at $r$, given chemical state $\nu$. For $r >0$ (corresponding to $x_0 > x_1$ when $\nu = 0$) the motor moves forward, and backwards otherwise. By the homodimer symmetry, ${\rm P}_{\text{sw}}(r|0) = {\rm P}_{\text{sw}}(-r|1)$.

By ergodicity, the trajectory average $\overline{\Delta r}$ can be expressed in terms of ensemble averages over ${\rm P}_{\text{sw}}(r|\nu)$ as
\begin{equation}\label{eq:mean-dis-general}
    \overline{\Delta r} = \int_{-\infty}^{\infty} {\rm d}r\ r \ [{\rm P}_{\text{sw}}(r|0) - {\rm P}_{\text{sw}}(r|1)] \ .
\end{equation}
The term within brackets is proportional to the steady-state probability flux appearing in the Fokker--Planck equations~\eqref{eq:PDE-marginal}. Equation~\eqref{eq:mean-dis-general} can therefore be interpreted as the average change of the relative displacement between transitions.

Similarly, the mean duration $\overline{\Delta t}$ can be expressed as the ensemble average of the inverse transition rates,  
\begin{subequations}\label{eq:mean-time}
\begin{align}
    \overline{\Delta t} &= \int_{-\infty}^{\infty} {\rm d}r \ k(r)^{-1} \ P_{\text{sw}}(r|0)
    \\
    &= \int_{-\infty}^{\infty} {\rm d}r \ k(-r)^{-1} \ P_{\text{sw}}(r|1)
    \\
    &= \left[\int_{-\infty}^{\infty}{\rm d}r \  k(r)P_{\infty}(r|0)\right]^{-1} \ .
\end{align}
\end{subequations}
The last line is the mean escape rate from state $\nu = 0$ which, by homodimer symmetry, equals that for $\nu = 1$. Thus, the mean duration $\overline{\Delta t}$ equals the inverse of the mean escape rate.

\subsection{Thermodynamic flows}\label{sec:thermo}
Transport motors, like many molecular machines, function as engines that transduce free energy between coupled subsystems. In our model, these subsystems are the mechanical ($X$) and chemical ($\nu$) degrees of freedom. The chemical dynamics for $\nu$, together with the Fokker–Planck equations~\eqref{eq:PDE-marginal} for the distributions of $r$, give rise to distinct thermodynamic flows constrained by ensemble-level subsystem-specific first laws 
\begin{subequations}\label{eq:first-law}
\begin{align}
      \dot{Q}_{\text{sw}} +  \dot{W}_{\text{chem}} &= 0 \label{eq:first-law-chem}
      \\
      \dot{Q}_{\text{diff}}^{(i)} + \dot{W}_{\text{mech}}^{(i)}
      &= \dot{E}^{(i)}, \quad  i= \{0,1\} \ . \label{eq:first-law-mech}
\end{align}
\end{subequations}
In state $0$, $\dot{E}^{(0)}$ is the rate of change of 
the coupling energy between $x_0$ and $x_1$ due to the dynamics of $x_0$, which can also be interpreted as the work done by the $x_0$ head on the $x_1$ head~\cite{ehrich_energy_2023}. Since $x_1$ is fixed, $\dot{E}^{(0)}$ can be expressed as
\begin{equation}
    \dot{E}^{(0)} = \frac{1}{\overline{\Delta t}} \int_{-\infty}^{+\infty}{\rm d}r \ V(r) [{\rm P}_{\text{sw}}(r|0) - {\rm P}_{\text{sw}}(r|1)] \ .
\end{equation}
By virtue of the homodimer symmetry (see SM I) that dictates an even $V(r)$, we have that $\dot{E}^{(0)} = 0$, and by the steady-state condition $\dot{E}^{(1)} = 0$ as well, implying that at steady state no energy is exchanged on average between the motor heads. We highlight that this is a consequence of the homodimer symmetry and does not necessarily hold for heterodimeric motors~\cite{bensel_mechanochemistry_2020}, or for motors deviating from the hand-over-hand stepping mechanism (e.g., using an inchworm mechanism~\cite{hua_distinguishing_2002,ciudad_dynamics_2006}). In such cases, the asymmetry between the two heads generically produces nonzero internal energy rates $\dot{E}^{(0)}$ and $\dot{E}^{(1)}$.

In SM I, we show that 
each thermodynamic flow is equal for the two motor heads, again by virtue of the homodimer symmetry. Therefore, we henceforth drop the superscript $i$ and (without loss of generality) consider all thermodynamic flows for $\nu = 0$.

The subsystem-specific first law~\eqref{eq:first-law-chem} does not include a rate of change of internal energy. By construction, in our model the switches do not alter the system's energy landscape, but rather restrict how the motor explores the landscape (i.e., with one head coordinate fixed at a time). This contrasts with traditional flashing-ratchet models, such as those used to describe the dynamics of $\text{F}_1$-ATPase~\cite{wareham_multiparameter_2025},  where each chemical state maps to a distinct energy profile. Consequently, in these models chemical transitions inject energy into the mechanical degrees of freedom via the instantaneous potential-energy differences between these landscapes. As we discuss later, it is the absence, or subdominant role, of internal energy flows that characterizes our model as an information engine. 

The total heat flow $\dot{Q} = \dot{Q}_{\rm diff} + \dot{Q}_{\text{sw}}$ has a contribution from the diffusive search of the motor heads~\cite{horowitz_multipartite_2015},
\begin{equation}\label{eq:q-diff}
    \dot{Q}_{\text{diff}} = \int_{-\infty}^{+\infty}{\rm d}r \ [V(r) + fr] \, \mathbb{L}_f^{(r)}[P_{\infty}(0,r)] \ ,
\end{equation}
and one due to chemical switches,
\begin{equation}\label{eq:q-switch}
    \beta\dot{Q}_{\text{sw}} = -\int_{-\infty}^{+\infty}{\rm d}r  \ P_{\infty}(0,r) \, k(r) \, \log \frac{k(r)}{k(-r)} \ .
\end{equation}
The total work flow $\dot{W} = \dot{W}_{\rm mech} + \dot{W}_{\rm chem}$ has a chemical contribution from the driving force due to the chemical free energy $\Delta \mu$,
\begin{equation}\label{eq:chem-work-general}
    \dot{W}_{\text{chem}} = \Delta \mu \int_{-\infty}^{+\infty}{\rm d}r \ h(r)\ k(r)\, P_{\infty}(0,r) \ ,
\end{equation}
and a mechanical contribution from the external force $f$,
\begin{equation}\label{eq:w-mech-general}
    \dot{W}_{\text{mech}} = -2f \int_{-\infty}^{+\infty}{\rm d}r \ r \ k(r)\, P_{\infty}(0,r) \ ,
\end{equation}
which can be further simplified to (see SM II)
\begin{equation}\label{eq:work-mech-force}
    \dot{W}_{\text{mech}} = -f \mathcal{V} \ .
\end{equation}
That is, if the directed motion opposes the constant force $f$ then $\dot{W}_{\text{mech}} < 0$, and hence $\dot{Q}_{\text{diff}} > 0$. The second law requires that the overall work flow $\dot{W}$ is positive and hence the overall heat flow $\dot{Q}$ is negative (otherwise the system would turn heat from a single thermal bath into work).

We define the thermodynamic efficiency as the ratio of the mechanical work output to the chemical work input~\cite{leighton_flow_2025},
\begin{equation}\label{eq:efficiency-td}
    \eta_{\text{TD}} \equiv \frac{-\dot{W}_{\text{mech}}}{\dot{W}_{\text{chem}}} \ .
\end{equation}
This metric quantifies how effectively the chemical free energy from ATP hydrolysis is converted into mechanical work against the cargo force, reporting on the proximity to the reversible limit, where energy transduction occurs entirely without dissipation. $\eta_{\text{TD}}$ of $0.4 - 0.6$ have been reported for kinesin~\cite{goychuk_anomalous_2015} and myosin~\cite{kjelstrup_mesoscopic_2013}, indicating that a significant fraction of the chemical free energy is dissipated as heat, and that these motors operate far from the reversible limit.

\begin{figure}
  \centering  \includegraphics[width=0.29\textwidth]{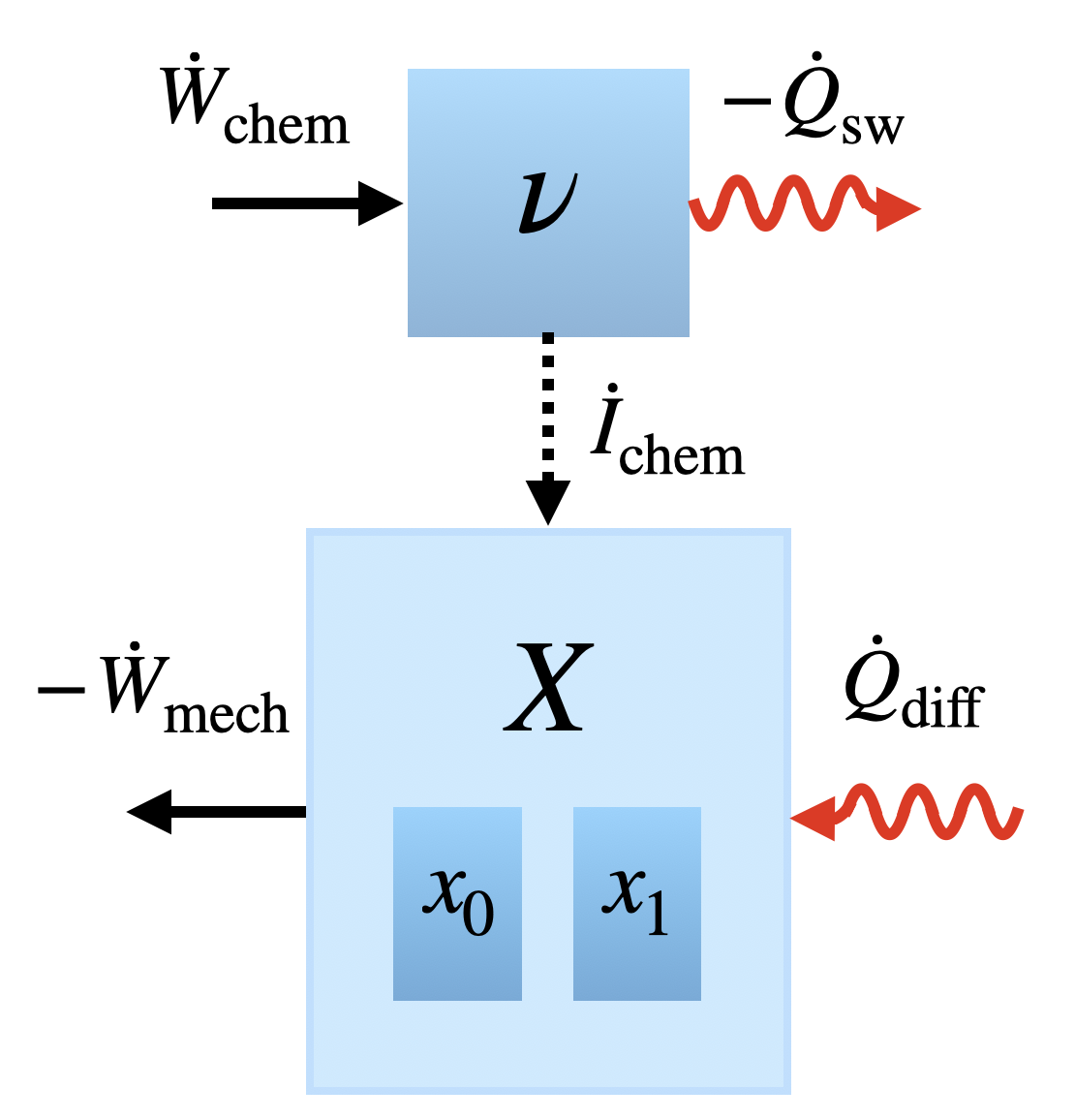}
  \caption{Operation of the switching-mobility motor as an information engine, illustrating the thermodynamic flows into and out of the mechanical degrees of freedom $X = (x_0,x_1)$ and chemical degree of freedom $\nu$.}
  \label{fig:info-engine_sketch}
\end{figure}

We have so far discussed the energy flows. In analogy with Eq.~\eqref{eq:first-law}, the global second law is
\begin{equation}\label{eq:second-law}
    \dot{\Sigma}_{\text{mech}} + \dot{\Sigma}_{\text{chem}} \geq 0 \ ,
\end{equation}
where $\dot{\Sigma}_{\text{mech}}$ and $\dot{\Sigma}_{\text{chem}}$ are the respective steady-state entropy production rates associated with the mechanical (Fokker--Planck) and chemical (master-equation) dynamics 
\cite{ehrich_energy_2023,horowitz_multipartite_2015}:
\begin{subequations}
    \begin{align}
        \dot{\Sigma}_{\text{mech}} &= D\int_{-\infty}^{+\infty}{\rm d}r \,
        \frac{\left[\beta ({\rm d}_r V + f)P_{\infty}(0,r) + {\rm d}_r P_{\infty}(0,r)\right]^2}{P_{\infty}(0,r)} \ , 
        \\
        \dot{\Sigma}_{\text{chem}} &= \int_{-\infty}^{+\infty}{\rm d}r \,
        k(r)\, P_{\infty}(0,r)
        \log \frac{k(r)P_{\infty}(0,r)}{k(-r)P_{\infty}(0,-r)} \ .
    \end{align}
\end{subequations}
Each contribution can be further decomposed as
\begin{subequations}\label{eq:decomp-entropy} 
\begin{align}
        \dot{\Sigma}_{\text{mech}} &= \dot{S}_{\text{mech}} - \beta \dot{Q}_{\text{diff}} \ ,
        \\
        \dot{\Sigma}_{\text{chem}} &= \dot{S}_{\text{chem}} - \beta\dot{Q}_{\text{sw}} \ . \label{subeq:second-law-chemical}
    \end{align}
\end{subequations}
The first right-hand-side terms are the respective rates of change of the system's Shannon entropy due to the mechanical and chemical dynamics,
\begin{subequations}\label{eq:system-entropy}
    \begin{align}
        \dot{S}_{\text{mech}} &= - \int_{-\infty}^{+\infty}{\rm d}r\,
        \mathbb{L}_f^{(r)}[P_{\infty}(0,r)]
        \log P_{\infty}(0,r) \ ,
        \\
        \dot{S}_{\text{chem}} &= \int_{-\infty}^{+\infty}{\rm d}r \,
        k(r)\, P_{\infty}(0,r)
        \log \frac{P_{\infty}(0,r)}{P_{\infty}(0,-r)} \ .
    \end{align}
\end{subequations}
At steady state, the total system entropy is constant, implying
\begin{equation}
    \dot{S}_{\text{mech}} = -\dot{S}_{\text{chem}} \ .
\end{equation}
The second right-hand-side terms in Eqs.~\eqref{eq:decomp-entropy} are the heat flows~\eqref{eq:q-diff} and \eqref{eq:q-switch}, equaling the environmental entropy production rates. The marginal entropies $S[X]$ and $S[\nu]$ are related to the joint entropy $S[X,\nu]$ through
\begin{equation}
    S[X] + S[\nu] = S[X,\nu] + I[X,\nu] \, ,
\end{equation}
for mutual information $I[X,\nu]$ between $X$ and $\nu$. At steady state, the marginal entropies do not depend on time, so that
\begin{equation}\label{eq:info-entropy-rel}
     \dot{I}[X,\nu] = -  \dot{S}[X,\nu] \, .
\end{equation}
In a similar manner to the Shannon entropy, we decompose the information flows into mechanical and chemical contributions. At steady state, Eq.~\eqref{eq:info-entropy-rel} then allows us to write
\begin{subequations}
    \begin{align}
        \dot{I}_{\text{mech}} &= -\dot{S}_{\text{mech}} \, ,\\
        \dot{I}_{\text{chem}} &= -\dot{S}_{\text{chem}} \, .
    \end{align}
\end{subequations}
The information flows $\dot{I}_{\text{mech}}$ and $\dot{I}_{\text{chem}}$ quantify the change in mutual information between $X$ and $\nu$ arising from the respective dynamics of $X$ and of $\nu$. In the absence of energy flow between the chemical and mechanical degrees of freedom, the motor solely transduces free energy between $\nu$ and $X$ in the form of information flow. The correlations between $\nu$ and $X$ are exploited to extract useful energy from the thermal environment via the ingoing diffusive heat flow $\dot{Q}_{\rm diff}$, while the chemical heat flow $\dot{Q}_{\rm sw}$ is out of the system. Figure~\ref{fig:info-engine_sketch} illustrates the thermodynamic flows underlying the operation of the motor as an information engine.

The bipartite structure further relates this free-energy transduction to the input and output work flows through the inequalities~\cite{leighton_flow_2025}
\begin{equation}\label{eq:inequality}
    \beta \dot{W}_{\text{chem}} \geq \dot{I}_{\text{chem}} \geq -\beta \dot{W}_{\text{mech}} \, .
\end{equation}
The information flow $\dot{I}_{\text{chem}}$ acts as a thermodynamic bottleneck, mediating the transduction between chemical work input and mechanical work output. This observation motivates the decomposition of the total thermodynamic efficiency, $\eta_{\text{TD}} = \eta_{\rm chem} \eta_{\rm mech}$, into separate subsystem free-energy transduction efficiencies for the chemical and mechanical degrees of freedom~\cite{leighton_flow_2025,leighton_information_2024}, 
\begin{subequations}\label{eq:subsystem-effs}
    \begin{align} 
        \eta_{\rm chem} &\equiv \frac{\dot{I}_{\text{chem}}}{\beta \dot{W}_{\text{chem}}} \leq 1 \, ,
        \\
        \eta_{\rm mech} &\equiv \frac{-\beta \dot{W}_{\text{mech}}}{\dot{I}_{\text{chem}}} \leq 1 \, .
    \end{align}
\end{subequations}

\section{Localized switching}\label{sec:fixed-step}
Certain transport motors, such as kinesin-1 and myosin-V, move along cytoskeletal filaments in a highly regular, stepwise manner. Kinesin-1 typically advances in $\sim$8-nm increments along microtubules~\cite{coy_kinesin_1999}, whereas myosin-V takes $\sim$36-nm steps along actin filaments~\cite{komori_myosin-v_2007}. This precise stepping behavior reflects a tight coupling between the structural periodicity of the filament and the motor’s mechanochemical cycle (that relates stepping mechanics and chemical free-energy consumption), thereby ensuring efficient and directed intracellular transport.

In this section, for analytical insight we spatially localize (to $r = \pm a$) the switching rate in the absence of chemical driving~\eqref{eq:transition-rates},
\begin{equation}\label{eq:rate-function}
    \Gamma (r) = k_0 \left[\delta(r-a) + \delta(r+a) \right] \ ,
\end{equation}
for bare switching rate $k_0$---giving bare switching timescale $\tau_0 \equiv k_0^{-1}$---and step length $a$. Using this, we compute the steady-state distribution $P_{\infty}(r|0)$ of relative displacements and then derive the performance metrics and thermodynamic flows. (SM III has details.)

The transition rate for localized switching reads
\begin{equation}\label{eq:rate-function-2}
    k^{\text{FS}}(r) \equiv k_0 \left[ e^{\beta\Delta \mu /2}\delta(r-a) + e^{-\beta\Delta \mu /2} \delta(r+a)\right] \ . 
\end{equation}
Although $\Gamma (r)$ is even in $r$, having equal propensity for forward and backward switches, the chemical driving $\Delta \mu$ breaks the spatial symmetry.

\subsection{Mean motor velocity}

The steady-state probability for the relative displacement at the next switch~\eqref{eq:distribution-r-displ} simplifies to
\begin{equation}\label{eq:rho-dist-fixed}
    {\rm P}_{\text{sw}}(r|0) = p({\rm f}) \ \delta(r-a) + p({\rm b}) \ \delta(r+a) \ ,
\end{equation}
with $p({\rm f})$ and $p({\rm b})$ the respective steady-state probabilities for forward and backward switches, satisfying $p({\rm f}) + p({\rm b}) = 1$:
\begin{subequations}\label{eq:ss-probs}
    \begin{align}
        p({\rm f}) &= \frac{k_+}{k_+ + k_-} \  \label{subeq:ss-probs-1}
        \\
        p({\rm b}) &= \frac{k_-}{k_++k_-} \ , \label{subeq:ss-probs-2}
    \end{align}
\end{subequations}
for respective steady-state rates $k_+$ and $k_-$ of the forward and backward switches,
\begin{subequations}\label{eq:ss-rates}
    \begin{align}
        k_+ &\equiv k_0 \, e^{\beta \Delta \mu / 2}P_{\infty}(a|0)
        \\
        k_- &\equiv k_0 \, e^{-\beta \Delta \mu / 2}P_{\infty}(-a|0) \ .
    \end{align}
\end{subequations}

All performance metrics can be expressed in terms of these rates. The mean displacement per cycle is
\begin{subequations}\label{eq:mean-dis-fs}
\begin{align}
    \overline{\Delta r} &= 2a \, [p({\rm f}) - p({\rm b})] \label{subeq:bias}
    \\
    &= 2a \, \frac{k_+-k_-}{k_++k_-} \ ,
\end{align}
\end{subequations}
and the mean duration of a head step is 
\begin{equation}\label{eq:mean-duration-fs}
    \overline{\Delta t} =  \frac{1}{k_++k_-} \ .
\end{equation}

Equations~\eqref{subeq:bias} and \eqref{eq:mean-duration-fs} offer a transparent physical interpretation. The mean displacement $\overline{\Delta r}$ of a single motor head equals the step length $2a$ multiplied by the net bias between forward and backward steps. Meanwhile, the mean duration of a head step is just the inverse of the sum of the two steady-state switching rates. Combining Eqs.~\eqref{eq:mean-vel-phys}, \eqref{eq:mean-dis-fs}, and \eqref{eq:mean-duration-fs} yields the mean motor velocity,
\begin{equation}\label{eq:velocity-fs}
    \mathcal{V} = 
    \frac{(e^{\beta\Delta \mathcal{F}}-1) \, a}
    {e^{\beta\Delta \mathcal{F}}(\tDiff^{-a\to a} + \tau_{\rm res}^a)+ \left(\tDiff^{a\to -a} + \tSW^{-a}\right)} \ ,
\end{equation}
for total free energy per forward switch
\begin{equation}\label{eq:free-energy-total}
    \Delta \mathcal{F} \equiv \Delta \mu - 2 f a \ ,
\end{equation}
diffusive mean first-passage times from $r = -a$ to $r = a$ and vice versa under the action of the total potential $V(r) + fr$, 
\begin{subequations}\label{eq:tmfp}
    \begin{align}
    \tDiff^{-a\to a}& \equiv \frac{1}{D}\int_{-a}^{a}{\rm d}r \ e^{\beta [V(r)+fr]} \int_{-\infty}^{r}{\rm d}r' \ e^{-\beta [V(r') + fr']} \label{subeq-dtmf-1}
    \\
    \tDiff^{a\to -a}&\equiv \frac{1}{D}\int_{-a}^{a}{\rm d}r \ e^{\beta [V(r)+fr]} \int_{r}^{\infty}{\rm d}r' \ e^{-\beta [V(r')+fr']} \ ,
    \end{align}
\end{subequations}
and mean residence times around the partially absorbing boundaries at $r = a$ and $r = -a$,
\begin{subequations}\label{eq:res-times}
    \begin{align}
        \tau_{\rm res}^{a} &\equiv \frac{e^{\beta [V(a)+fa]}}{2k_0 \, e^{\beta \Delta \mu /2}} \int_{-\infty}^{+\infty}{\rm d}r \ e^{-\beta [V(r)+fr]}
        \\
        \tau_{\rm res}^{-a} &\equiv \frac{e^{\beta [V(a)-fa]}}{2k_0 \, e^{-\beta \Delta \mu /2}} \int_{-\infty}^{+\infty}{\rm d}r \ e^{-\beta [V(r)+fr]} \ .
    \end{align}
\end{subequations}
The times in Eqs.~\eqref{eq:tmfp} and \eqref{eq:res-times} are obtained from the adjoint Fokker--Planck equation associated with Eq.~\eqref{eq:PDE-marginal}, subject to different boundary conditions~\cite{kampen_stochastic_1992}. They characterize the distinct stages involved in completing a motor step. The diffusive times $\tau_{\rm diff}^{-a \to a}$ and $\tau_{\rm diff}^{a \to -a}$ quantify the mean time for the motor head to travel between neighboring binding sites under the total potential $V(r)+fr$. In contrast, the residence times $\tau_{\rm res}^{-a}$ and $\tau_{\rm res}^{a}$ measure the average time spent near a binding site before a switch. These residence times can also be interpreted as the respective inverses of the equilibrium rates for forward and backward switches.

Equation~\eqref{eq:velocity-fs} reproduces the experimentally observed behavior of single kinesin motors~\cite{svoboda_force_1994}. In particular, the factor $(e^{\beta \Delta \mathcal{F}} - 1)$ can be interpreted as the thermodynamic driving force, which depends on the balance between the chemical free energy $\Delta \mu$ (itself controlled by the ATP concentration) and the mechanical work $2fa$ required to complete a forward step. For chemical driving strongly exceeding mechanical driving, the velocity saturates at a maximum value
\begin{subequations}\label{eq:maximal-vel}
    \begin{align}
    \mathcal{V}_{\text{max}} &\equiv \lim_{\beta\Delta \mathcal{F}\to \infty} \mathcal{V}
    \\
    &=\frac{a}{\tDiff^{-a\to a}} \ .
    \end{align}
\end{subequations}
This maximal velocity is determined by the system's mechanical properties, such as the shape of the linker potential $V(r)$, the cargo force $f$, and the step size $a$. This saturation is a common feature in models of molecular motors, reflecting that there are intrinsic limits to how fast a motor can move, regardless of the available chemical free energy~\cite{pietzonka_universal_2016,svoboda_force_1994}.

The free energy $\Delta \mathcal{F}$ per completed step constitutes the thermodynamic driving force for the motion: its sign determines the direction of transport. In particular, for $\Delta \mathcal{F} > 0$, the motion is biased forward, with $p({\rm f}) > p({\rm b})$. The mean velocity vanishes ($\mathcal{V}=0$) at the stall force
\begin{equation}\label{eq:stall-force}
    f_{\rm s} = \frac{\Delta \mu}{2a} \ ,
\end{equation}
for which $\Delta \mathcal{F} = 0$. The fact that the stall force corresponds to the chemical free energy per step length is a hallmark of tight coupling~\cite{wang_stokes_2002}, i.e., a one-to-one correspondence between chemical switches and motor-head steps against the cargo force. The stall force is the value for which the mechanical work $f \times 2a$ performed over a full step of length $2a$ against the constant force $f$ equals the chemical free energy $\Delta \mu$ supplied per forward switch. For $f > f_{\rm s}$, the chemical driving is insufficient to sustain forward motion, and the motor moves backward on average.

\begin{figure*}[t]
\centering
\includegraphics[width=1.0\textwidth]{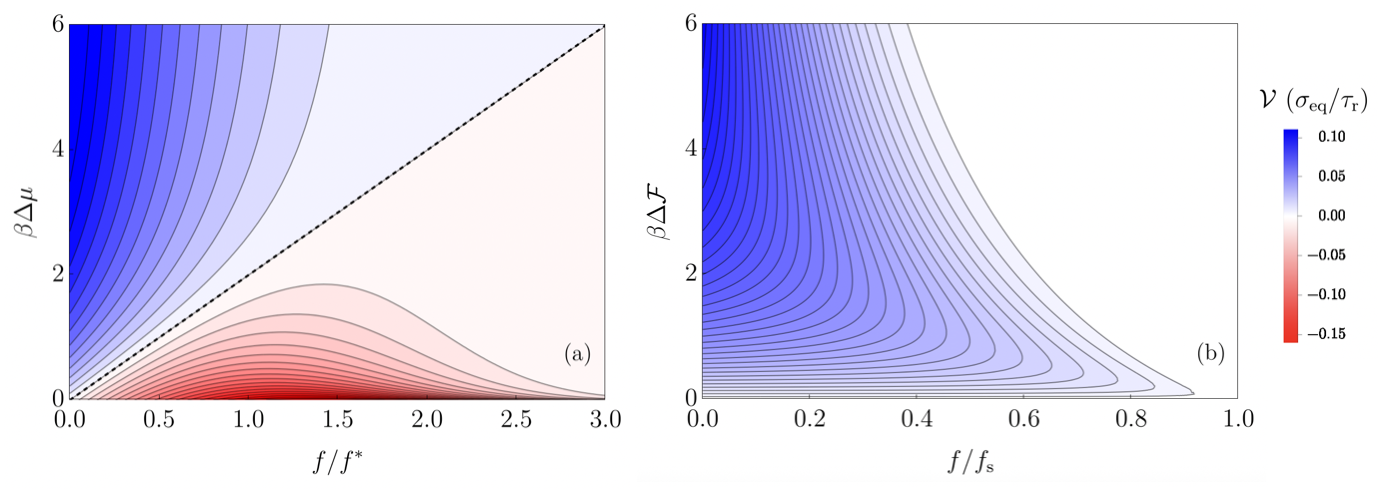}
\caption{Heatmaps of the mean motor velocity $\mathcal{V}$. (a) $\mathcal{V}$ as a function of the chemical free energy $\beta \Delta \mu$ and the ratio $f/f^*$, with $f^*$ the characteristic force scale. The black dashed line indicates the stall regime ($\Delta \mu = 2fa$), where the velocity vanishes ($\mathcal{V} = 0$). (b) $\mathcal{V}$ as a function of the total free energy $\Delta \mathcal{F}$ and the force ratio $f/f_{\rm s}$ (for stall force $f_{\rm s}$~\eqref{eq:stall-force}), which also equals the thermodynamic efficiency $\eta_{\rm TD}$. Throughout, we consider bare switching rate $k_0 = \tR^{-1}$, step length $a = \sigma_{\rm eq}$, and harmonic linker potential $V(r) = \frac{1}{2}\kappa r^2$.}
\label{fig:vel-maps}
\end{figure*}

Figure~\ref{fig:vel-maps}a illustrates the mean motor velocity $\mathcal{V}$ as a function of the chemical driving $\Delta \mu$ and cargo force $f$, with fixed kinetic parameters $k_0 = \tau_{\rm r}^{-1}$ and $a = \sigma_{\rm eq}$. The stall regime ($f = f_{\rm s}$) partitions the parameter space into two distinct regions: one where the motor goes forward (blue) and one where it goes backward (red).

The mean velocity is markedly asymmetric across the stall force. In the forward regime $(\mathcal{V}> 0)$, the velocity is maximal at zero cargo force. With increasing chemical free energy $\Delta \mu$, the contour lines become vertical, indicating that for a given $f$ the velocity saturates to its maximum value $\mathcal{V}_{\text{max}}$ [from Eq.~\eqref{eq:maximal-vel}]. Conversely, in the backward regime ($\mathcal{V} < 0$), the motor behaves quite differently. For fixed $\Delta \mu$, the velocity magnitude is maximal for nonzero $f$, and it decays to zero for excessively high cargo forces.

Figure~\ref{fig:vel-maps}b depicts the mean motor velocity $\mathcal{V}$ for forward motion as a function of the total free energy $\Delta \mathcal{F}$ and the force ratio $f/f_{\rm s}$. Consistent with the trends in Fig.~\ref{fig:vel-maps}a, for a fixed force the velocity saturates at high $\Delta \mathcal{F}$, while the velocity monotonically vanishes while approaching stall ($f \to f_{\rm s}$). Notably, at fixed $f/f_{\rm s}$ a vertical line crosses the same contour line twice, i.e., the same velocity $\mathcal{V}$ can be achieved with different values of $\Delta \mathcal{F}$. This behavior highlights that motor transport is not solely governed by the thermodynamic imbalance between chemical and mechanical forces, but is critically constrained by internal mechanics, specifically the mechanics of stepping between adjacent binding sites. Regardless of the magnitude of the chemical free energy, switching remains impossible if the mechanical properties of the linker potential and cargo force do not permit the motor head to traverse the step distance $2a$. We further discuss this behavior in Sec.~\ref{sec:timescale}.

\begin{figure} 
  \centering  \includegraphics[width=0.45\textwidth]{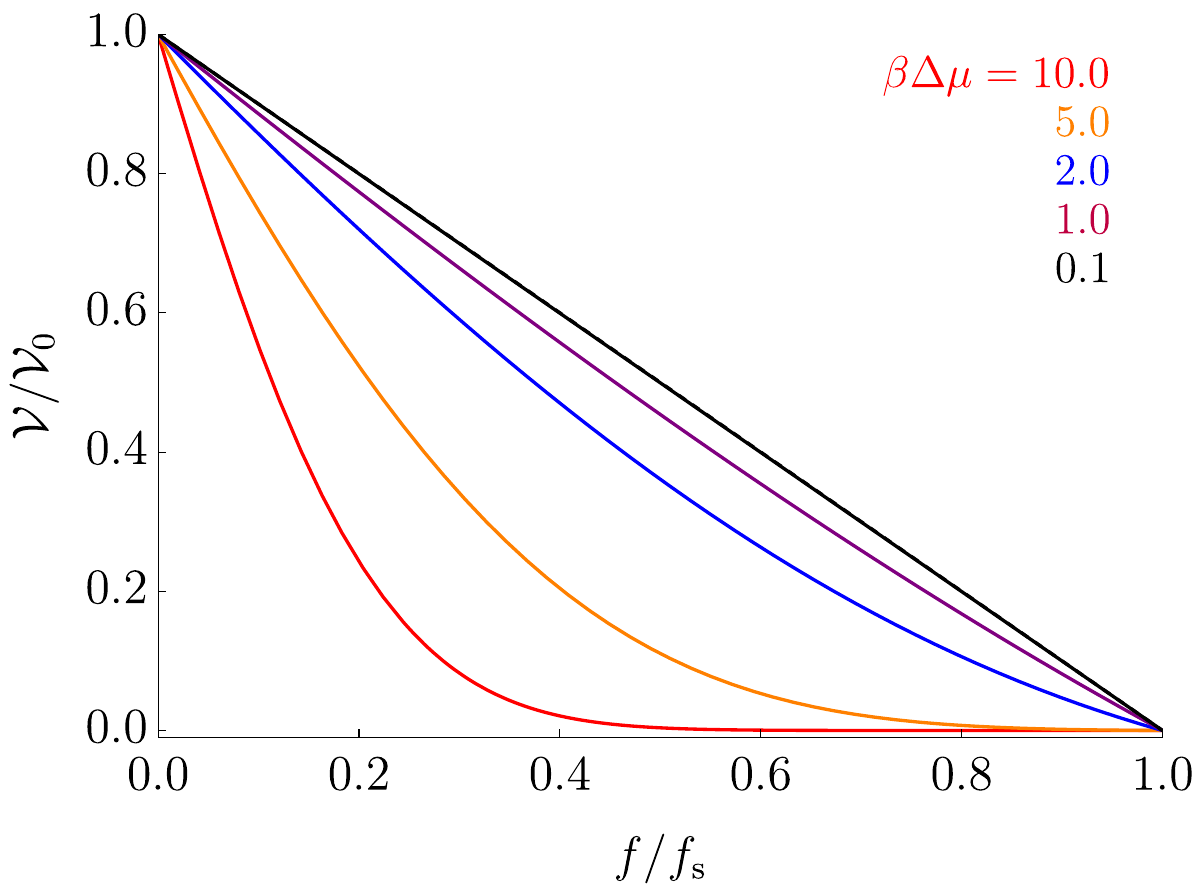}
  \caption{Ratio $\mathcal{V}/\mathcal{V}_0$ of velocity $\mathcal{V}$ to velocity $\mathcal{V}_0$ at zero cargo force, as a function of the force ratio $f/f_{\rm s}$, for different chemical free energies $\beta \Delta \mu$, with bare switching rate $k_0 = \tR^{-1}$, step length $a = \sigma_{\rm eq}$, and harmonic confinement $V(r) = \frac{1}{2}\kappa r^2$.}
  \label{fig:vel-relations}
\end{figure}

For kinesin motors with external resisting force smaller than the stall force, experiments have shown that the mean motor velocity is approximately linear with the applied force~\cite{svoboda_force_1994,hunt_force_1994,leighton_performance_2022},
\begin{equation}\label{eq:linear-vel}
    \mathcal{V} = \mathcal{V}_0 \left(1 - \frac{f}{f_{\rm s}}\right) \,.
\end{equation}
Here $\mathcal{V}_0$ is the maximal velocity attained at zero cargo force, holding fixed all other model parameters. 

To probe the limits of this relation within our model, Fig.~\ref{fig:vel-relations} shows the ratio $\mathcal{V}/\mathcal{V}_0$ as a function of the force ratio $f/f_{\rm s}$. The linear behavior predicted by Eq.~\eqref{eq:linear-vel} is only obtained for small total free energy $\Delta \mathcal{F}$ (i.e., near equilibrium). In this regime, the maximal velocity $\mathcal{V}_0$ from Eq.~\eqref{eq:linear-vel} becomes
\begin{equation}\label{eq:vel-0}
    \mathcal{V}_0 = \frac{a \beta\Delta \mu}{\tDiff^{-a\to a}  + \tDiff^{a\to -a} + 2\tSW^{\rm eq}} \ ,
\end{equation}
for the equilibrium total mean residence time
\begin{equation}
    \tSW^{\rm eq} \equiv \frac{e^{\beta V(a)}}{2k_0 } \int_{-\infty}^{+\infty}{\rm d}r \ e^{-\beta [V(r)+fr]} \ .
\end{equation}
Note that $2\tSW^{\rm eq}$ is the sum of the mean residence times in Eq.~\eqref{eq:res-times} under stall conditions ($\Delta \mathcal{F}=0$). From Eq.~\eqref{eq:vel-0}, it is evident that increasing $\Delta \mu$ enhances the maximal velocity. For $\Delta \mu \gg 2fa$, as predicted by Eq.~\eqref{eq:maximal-vel}, the velocity saturates and exhibits a clear nonlinear dependence, arising from the dependence in Eq.~\eqref{subeq-dtmf-1} of the diffusive mean first-passage time $\tau_{\rm diff}^{-a \to a}$ on the cargo force $f$.

\subsection{Thermodynamic flows}
We focus now on the thermodynamic flows from Sec.~\ref{sec:thermo}. The chemical work rate for localized switching simplifies to
\begin{equation}\label{eq:chem-work-fs}
    \dot{W}_{\text{chem}} = \frac{\Delta\mu}{2a}\, \mathcal{V} \ .
\end{equation}
Using the chemical-subsystem first law~\eqref{eq:first-law-chem}, together with the simplified expression for the mechanical work flow~\eqref{eq:work-mech-force}, the total work and heat flows are
\begin{subequations}\label{eq:thd-flows-force}
    \begin{align}
        \dot{W} &= \left( \frac{\Delta\mu}{2a} - f \right) \mathcal{V} \ge 0 \  \\
        \dot{Q} &= -\left( \frac{\Delta\mu}{2a} - f \right) \mathcal{V} \le 0 \ .
    \end{align}
\end{subequations}
The expression in parentheses is precisely $f_{\rm s}-f$. $\mathcal{V}$ has the same sign as $f_{\rm s}-f$, such that the sign of each thermodynamic flow in Eq.~\eqref{eq:thd-flows-force} is independent of $\Delta \mu$ and $f$. For $f = f_{\rm s}$, $\mathcal{V}=0$ and $\dot{W} = \dot{Q} = 0$, indicating no net motion at equilibrium.

The chemical work and the mechanical work are proportional to the velocity, thus the thermodynamic efficiency from Eq.~\eqref{eq:efficiency-td} simplifies to
\begin{subequations}\label{eq:efficiency-fs}
    \begin{align}
    \eta_{\text{TD}} &= \frac{2af}{\Delta\mu}
    \\
    &= \frac{f}{f_{\rm s}} \ .
    \end{align}
\end{subequations}
This linear relation between $\eta_{\text{TD}}$ and the applied force $f$ is another hallmark of tight coupling. The motor operates at maximum efficiency ($\eta_{\text{TD}} = 1$) precisely at the stall force, where all the chemical free energy is converted into mechanical work without any dissipation~\cite{wang_stokes_2002}. $\eta_{\text{TD}}$ is only physically meaningful for the intended operation when the motor moves forward against the cargo force, i.e., for (without loss of generality) $\Delta \mu \geq 0$ and $0 \leq f \leq f_{\rm s}$.

The total entropy production rates for localized switching reduce to
\begin{subequations}
    \begin{align}
        \dot{\Sigma}_{\text{mech}} &= \frac{\Sigma}{2a} \mathcal{V}  \label{subeq:}
        \\
        \dot{\Sigma}_{\text{chem}} &= -\beta\dot{Q} - \frac{\Sigma}{2a} \mathcal{V} \ , \label{subeq:total-entr-prod-chem}
    \end{align}
\end{subequations}
where
\begin{equation}\label{eq:entropy-FP}
    \Sigma \equiv \beta \Delta \mathcal{F} - \log \frac{k_+}{k_-}
\end{equation}
is the total entropy change associated with the Fokker--Planck dynamics over a single step. Thus $\Sigma$ has the same sign as $\Delta \mathcal{F}$.

It is straightforward to verify that the global second law~\eqref{eq:second-law} is satisfied,
\begin{equation}
    \dot{\Sigma}_{\text{mech}} + \dot{\Sigma}_{\text{chem}} = -\beta\dot{Q} \geq 0 \,.
\end{equation}
Combining Eqs.~\eqref{subeq:second-law-chemical} and \eqref{subeq:total-entr-prod-chem} yields the rate of change of system entropy associated with the chemical dynamics,
\begin{equation}
    \dot{S}_{\text{chem}} = - \frac{\Sigma}{2a} \mathcal{V} - \beta\dot{Q}_{\text{diff}} \ ,
\end{equation}
from which the chemical information flow is
\begin{equation}\label{eq:chem-info-flow}
    \dot{I}_{\text{chem}} = \frac{\Sigma}{2a} \mathcal{V} + \beta\dot{Q}_{\text{diff}} \ .
\end{equation}

The subsystem efficiencies from Eq.~\eqref{eq:subsystem-effs} are simply
\begin{subequations}
    \begin{align}
        \eta_{\rm chem} &= \frac{\Sigma + 2\beta fa}{\beta \Delta \mu} 
        \\
        \eta_{\rm mech} &= \frac{2\beta fa}{\Sigma + 2\beta fa} \ .
    \end{align}
\end{subequations}

\begin{figure}
  \centering  \includegraphics[width=0.48\textwidth]{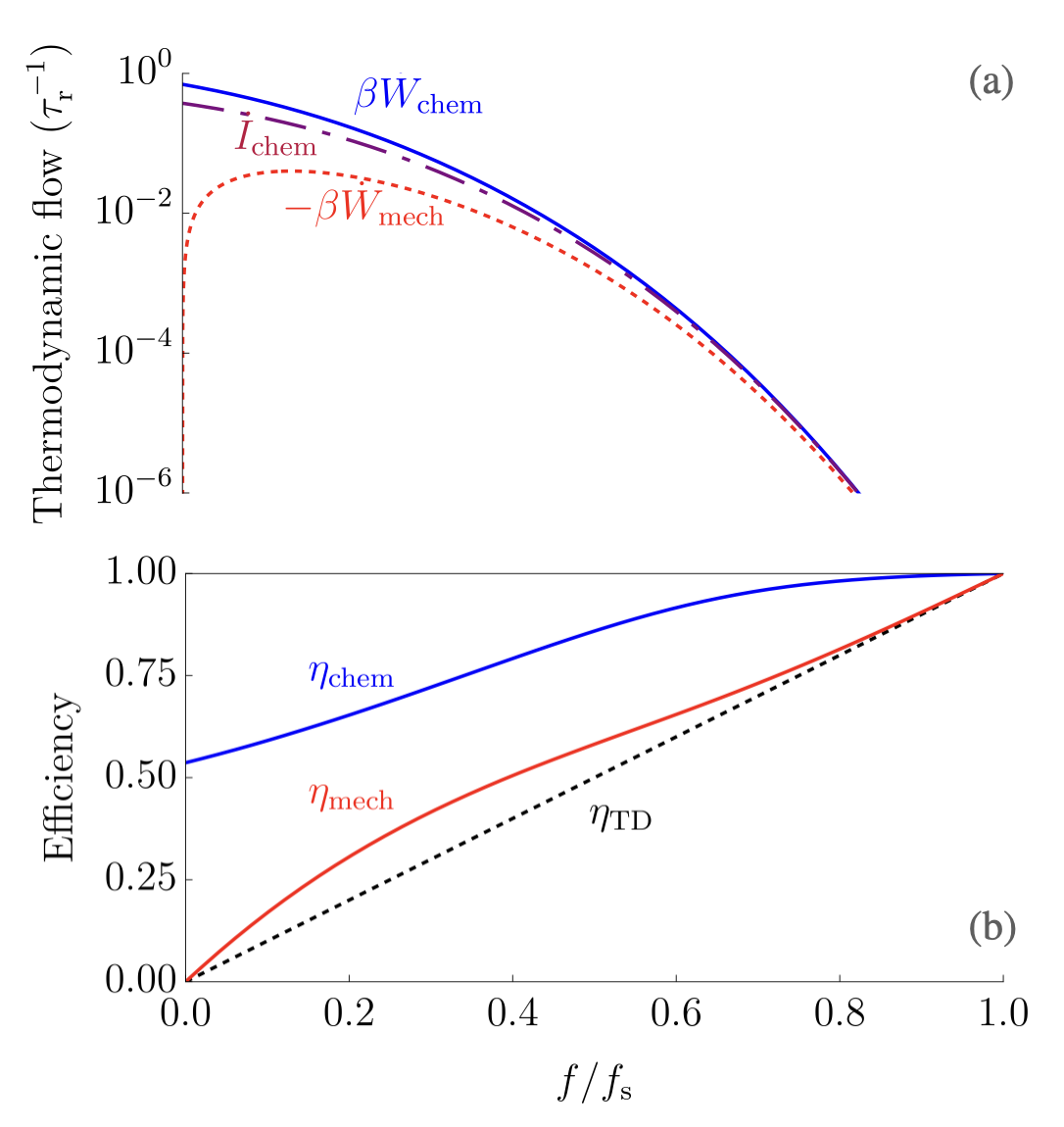}
  \caption{(a) Energy and information flows for localized switching. Chemical work rate $\beta\dot{W}_{\text{chem}}$ \eqref{eq:chem-work-fs}, mechanical work rate $-\beta\dot{W}_{\text{mech}}$ \eqref{eq:work-mech-force}, and chemical information flow $\dot{I}_{\text{chem}}$ \eqref{eq:chem-info-flow}, all as functions of the force ratio $f/f_{\rm s}$ (for stall force $f_{\rm s}$~\eqref{eq:stall-force}), with chemical free energy $\beta \Delta \mu = 10$, bare switching rate $k_0 = \tR^{-1}$, step length $a = \sigma_{\rm eq}$, and harmonic confinement $V(r) = \tfrac{1}{2}\kappa r^2$. (b) Overall efficiency $\eta_{\text{TD}}$ \eqref{eq:efficiency-td} and subsystem efficiencies $\eta_{\rm chem}$ and $\eta_{\rm mech}$ \eqref{eq:subsystem-effs}, satisfying $\eta_{\text{TD}} = \eta_{\rm chem}\eta_{\rm mech}$.}
  \label{fig:thd-flows}
\end{figure}

Figure~\ref{fig:thd-flows}a shows the work and information flows associated with the motor's operation as a function of the force ratio $f/f_{\rm s}$. As expected, all thermodynamic flows vanish in the stall regime $f = f_{\rm s}$. The mechanical work rate $-\beta\dot{W}_{\text{mech}}$ is nonmonotonic, as it vanishes in the absence of the cargo force. The thermodynamic inequalities in Eq.~\eqref{eq:inequality} are clearly satisfied, since the information flow always lies between the work flows. This is further corroborated in Figure~\ref{fig:thd-flows}b, showing that the thermodynamic efficiency $\eta_{\text{TD}}$ and subsystem efficiencies $\eta_{\rm chem}$ and $\eta_{\rm mech}$ remain below unity, and approach unity only at the stall force.

\subsection{Timescale separation and kinetic approach}\label{sec:timescale}
When modeling transport motors with localized switching, it is common to adopt the discrete stochastic framework used in traditional chemical kinetics. In this kinetic approach, the mechanical details are coarse-grained, with the relevant degrees of freedom represented as discrete states corresponding to the multiple mechanochemical states involved in motor stepping between binding sites, such as ATP binding or the release of ADP (adenosine diphosphate) and inorganic phosphate. Transitions between states are governed by rates that satisfy local detailed balance, which encodes the underlying energetics (the chemical free energy transduced during the corresponding biochemical transitions).

\begin{figure}
  \centering  \includegraphics[width=0.6\columnwidth,height=0.2\columnwidth]{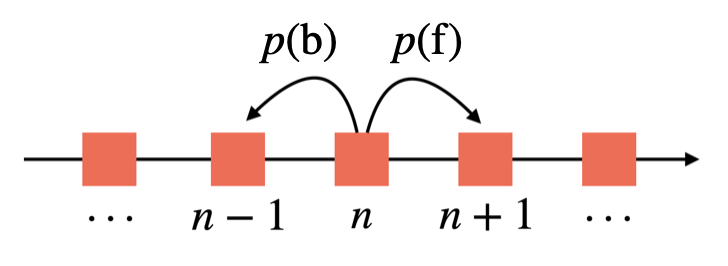}
  \caption{Schematic of the simplest kinetic approach for transport-motor dynamics. The motor jumps between discrete states corresponding to binding sites along the track.}
  \label{fig:kinetic-approach}
\end{figure}

The simplest kinetic approach for localized switching (see Fig.~\ref{fig:kinetic-approach}) is a single degree of freedom that hops between neighboring discrete states (labeled by $n$), corresponding to successive binding sites. Here the forward switch probabilities $p^{({\rm kin})}({\rm f})$ and backward switch probabilities $p^{({\rm kin})}({\rm b})$ satisfy the detailed-balance relation 
\begin{equation}\label{eq:kin-det-bal}
    \log \frac{p^{({\rm kin})}({\rm f})}{p^{({\rm kin})}({\rm b})} = \beta \Delta \mathcal{F} \,
\end{equation}
for the total free energy $\Delta \mathcal{F}$ per completed step~\eqref{eq:free-energy-total}. I.e., the sign of $(f_{\rm s}-f) \propto \Delta \mathcal{F}$ determines the direction of motion. Comparing this relation with the switch probabilities~\eqref{eq:ss-probs} obtained from the fundamental model,
\begin{equation}\label{eq:ldb-2}
    \log 
    \frac{p({\rm f})}{p({\rm b})}
    = \beta \Delta \mathcal{F} - \Sigma \ ,
\end{equation}
there is an additional contribution $\Sigma$ quantifying the total entropy change of the coarse-grained degrees of freedom~\eqref{eq:entropy-FP}. In our model, each motor head must first reach the target binding site through purely mechanical dynamics (set by the linker potential and the cargo force); only upon reaching this site does the coupling between chemical switches and mechanical motion become relevant, thereby introducing a bias in the mobility switching.

The scale of this entropic contribution depends on the interplay between the different times governing mechanical stepping and chemical switching. These include the diffusive times $\tau_{\rm diff}^{-a \to a}$ and $\tau_{\rm diff}^{a \to -a}$ from Eq.~\eqref{eq:tmfp} and the residence times $\tau_{\rm res}^a$ and $\tau_{\rm res}^{-a}$ that the motor head spends around each binding site before switching. Recovering Eq.~\eqref{eq:kin-det-bal} from Eq.~\eqref{eq:ldb-2} requires a strong separation of timescales (see SM III), 
\begin{equation}\label{eq:timescale-separation}
    {\rm max}(\tau_{\rm res}^a, \tau_{\rm res}^{-a}) \gg \tau_{\rm diff}^{a \to -a} + \tau_{\rm diff}^{-a \to a} \ .
\end{equation}
In this limit, the steady-state switching rates~\eqref{eq:ss-rates} are dominated by the residence times,
\begin{subequations}\label{eq:limiting-rates}
    \begin{align}
        k_+ &\sim \frac{1}{2\tau_{\rm res}^{a}}
        \\
        k_- &\sim \frac{1}{2\tau_{\rm res}^{-a}} \ .
    \end{align}
\end{subequations}

Substituting into Eq.~\eqref{eq:entropy-FP} gives
$\Sigma \to 0$. Physically, this limit corresponds to the chemical dynamics being much slower than the mechanics. Since chemical switches are rare, the mobile motor head has sufficient time to explore the available configurational space. As a result, the mechanical degrees of freedom relax to equilibrium between successive switches, causing the associated diffusive entropy production to vanish.

In the opposite limit,
\begin{equation}\label{eq:timescale-separation-2}
    {\rm max}(\tau_{\rm res}^a, \tau_{\rm res}^{-a}) \ll \tau_{\rm diff}^{a \to -a} + \tau_{\rm diff}^{-a \to a} \ ,
\end{equation}
the chemical dynamics is much faster than the mechanics, and the chemical degrees of freedom equilibrate. Thus, the total entropy production rate for the chemical dynamics vanishes, $\dot{\Sigma}_{\text{chem}} \to 0$, as does the bias between forward and backward switches,
\begin{equation}\label{eq:vanishing-bias}
    \frac{|p({\rm f})-p({\rm b})|}{\sqrt{p({\rm f})p({\rm b})}} \rightarrow 0 \ .
\end{equation}
As noted in previous works on flashing-ratchet models~\cite{makhnovskii_flashing_2004}, the kinetic description breaks down in the limit of fast chemical dynamics, as the time between switches becomes too short for a head to explore space and locate a new binding site; consequently, the mechanism responsible for directed motion fails. In contrast, in our model the mobility switches exclusively at localized positions, which prevents this issue. As a result, the motor attains a finite velocity, obtained in Eq.~\eqref{eq:velocity-fs} in the limit of vanishing residence times. Since both residence times in Eq.~\eqref{eq:res-times} are proportional to the bare switching timescale $\tau_0$, this limit corresponds to $\tau_0$ much shorter than the mechanical relaxation time $\tau_{\rm r}$,
\begin{equation}\label{eq:mean-vel-vanishing-res}
        \lim_{\tR/\tau_0 \to \infty}\mathcal{V}
        = \frac{(e^{\beta\Delta \mathcal{F}}-1)a}{e^{\beta\Delta \mathcal{F}}\tDiff^{-a\to a} + \tDiff^{a\to -a}} \, .
\end{equation}
Despite the coarse-grained description having asymptotically equal probabilities for forward and backward steps~\eqref{eq:vanishing-bias}, directed motion persists, controlled by the free energy $\Delta \mathcal{F}$. The emergence of a finite velocity from the combined limits of vanishing bias and increasing activity ($\tR/\tau_0 \to \infty$) has been reported in numerical studies of coupled rotaxane-based motors~\cite{gu_it_2026}.

Thermodynamically, the subsystem efficiencies defined in Eqs.~\eqref{eq:subsystem-effs} provide insight into these limiting regimes. In the fast chemical-driving limit ($\tR/\tau_0 \gg 1$), $\eta_{\rm chem} \to 1$ and $\eta_{\rm mech} \to \eta_{\text{TD}}$, indicating that the chemical degree of freedom $\nu$ efficiently transduces chemical work into information flow with negligible dissipation. In the opposite limit ($\tR/\tau_0 \ll 1$), $\eta_{\rm chem} \to \eta_{\text{TD}}$ and $\eta_{\rm mech} \to 1$, indicating that 
the information flow is perfectly converted into mechanical work, therefore any losses occur in converting chemical free energy into information flow.

\subsection{Dwell-time distributions}\label{sec:dwell-time-dis}
In single-molecule experiments, the stochastic trajectory of a transport motor can be directly tracked with high spatial and temporal resolution~\cite{higuchi_kinetics_1997,wirth_minflux_2023}. As the motor transitions between successive binding sites, its motion exhibits distinct metastable mechanical states associated with these sites, rather than appearing purely diffusive. Central quantities derived from such trajectories are the dwell-time distributions, which characterize the probability of the motor remaining in a given binding site for a certain time interval before taking either a forward or backward step. 

For traditional chemical kinetic models, the time a motor spends in a binding site is a Poisson process, and in particular does not depend on where it previously was. Therefore, the dwell-time distributions for both forward and backward steps are purely exponential functions with common decay rates~\cite{kolomeisky_molecular_2007}. However, upon coarse-graining models with continuous translational degrees of freedom, the dwell-time distributions are governed by the interplay of timescales associated with internal relaxation and barrier crossing in the free-energy landscape. When the barrier height significantly exceeds thermal energy (the Kramers regime), the system reaches local thermodynamic equilibrium within the well long before a rare escape event. This separation of timescales effectively 'erases' memory of where the motor came from, reducing the continuous diffusive crossing to a discrete Markovian jump process with a single-exponential dwell-time distribution~\cite{hanggi_reaction_1990}. When the energetic barriers are comparable to the thermal energy, the relevant degrees of freedom do not fully explore the well before switching. This yields non-exponential dwell times in the discrete coarse-graining, as the probability of switching depends on the motor's initial position on the landscape. In the more general scenario, it is this non-exponential behavior that encodes core information about the mechanochemistry of stepping.

In continuous-state descriptions, dwell times are determined by the probability fluxes leaving the current chemical state, which act as sink terms in the corresponding Fokker--Planck equation~\eqref{eq:fokker-planck}. Thus, we obtain dwell-time distributions by solving the closed system of modified Fokker-Planck equations with sinks~\cite{kampen_stochastic_1992}:
\begin{subequations}\label{eq:motor-kinetics-waiting-time}
\begin{align}
    \partial_t P^{(\text{in})}_t(0,r|\sigma) = 
    &\ \mathbb{L}_{f}^{(r)}[P^{(\text{in})}_t(0,r|\sigma)]
    - k^{(\text{FS})}(r) P^{(\text{in})}_t(0,r|\sigma) \nonumber
    \\
    &\ + k_0 e^{-\beta\Delta\mu/2}\,\delta(r-a)\, P^{(\text{in})}_t(1,r|\sigma) \label{subeq:dtd-fp-0}
    \\
    \partial_t P^{(\text{in})}_t(1,r|\sigma) = 
    &\ \mathbb{L}_{-f}^{(r)}[P^{(\text{in})}_t(1,r|\sigma)]
    - k^{(\text{FS})}(-r) P^{(\text{in})}_t(1,r|\sigma) \label{subeq:dtd-fp-1}
    \nonumber
    \\
    &\ + k_0 e^{\beta \Delta\mu/2}\,\delta(r-a)\, P^{(\text{in})}_t(0,r|\sigma) 
    \\
    {\rm d}_t p^{{(\text{out})}}_t(\sigma) =&\ k_0 e^{\beta \Delta \mu/2}P^{(\text{in})}_t(1,-a|\sigma) \nonumber
    \\
    &\ + k_0 e^{-\beta \Delta\mu/2}\,P^{(\text{in})}_t(0,-a|\sigma) \ , \label{subeq:dtd-fp-out}
\end{align}
\end{subequations}
for the probability $P^{(\text{in})}_t(\nu,r|\sigma)$ of finding the motor at chemical state $\nu$, with relative displacement $r$, at time $t$ within the current binding site, and the probability $p^{{(\text{out})}}_t(\sigma)$ that the motor has already left the current binding site at time $t$. Here, $\sigma \in \{0,1\}$ denotes the initial chemical state. These probabilities satisfy the normalization condition
\begin{equation}
    p_t^{(\text{out})}(\sigma) +
    \sum_{\nu = 0}^1\int_{-\infty}^{+\infty}{\rm d}r \ P_t^{(\text{in})}(\nu,r|\sigma) 
    = 1 \quad \forall t \ ,
\end{equation}
and are complemented with the initial conditions
\begin{subequations}\label{eq:initial-conditions}
\begin{align}
    P^{(\text{in})}_0(0,r|\sigma) &= (1-\sigma)\,\delta(r+a) \ \\
    P^{(\text{in})}_0(1,r|\sigma) &= \sigma\,\delta(r+a) 
    \\
    p_0^{(\text{out})}(\sigma) &= 0 \ .
\end{align}
\end{subequations}
Note that the only difference between Eqs.~(\ref{subeq:dtd-fp-0},\ref{subeq:dtd-fp-1}) and Eqs.~\eqref{eq:fokker-planck} is that the latter has source terms corresponding to transitions between states $0$ and $1$ at $r = -a$. Without loss of generality, at $r = -a$, a $0\to 1$ transition corresponds to a backward step (switching at $x_0 = x_1 - a$), whereas a $1\to 0$ transition corresponds to a forward step (switching at $x_1 = x_0 + a$). 

Regardless of the initial chemical state $sigma$, the solutions of Eq.\eqref{eq:motor-kinetics-waiting-time} transfer probability mass from $P_t^{(\text{in})}(\nu,r|\sigma)$ to $p_t^{(\text{out})}(\sigma)$ through the absorbing boundaries at $r = -a$ whenever the motor transitions to a new binding site. The probability fluxes at such boundaries are
\begin{subequations}\label{eq:dwell-time-dis}
\begin{align}
    j({\rm f}, t|\sigma) &\equiv k_0 \, e^{\beta\Delta\mu/2}\,P^{(\text{in})}_t(1,-a|\sigma) \ \\
    j({\rm b}, t|\sigma) &\equiv k_0 \, e^{-\beta\Delta\mu/2}\,P^{(\text{in})}_t(0,-a|\sigma) \ ,
\end{align}
\end{subequations}
and, for each $\sigma$, they satisfy the normalization condition
\begin{equation}\label{eq:dtd-norm}
    p({\rm f}|\sigma) +  p({\rm b}|\sigma) = 1 \ ,
\end{equation}
for the respective probabilities 
\begin{equation} 
    p(\Omega|\sigma) \equiv \int_0^{\infty}{\rm d}t \ j(\Omega,t|\sigma) \ ,
\end{equation}
for forward and backward stepping, i.e., $\Omega = \{{\rm f}, {\rm b}\}$, conditioned on the initial chemical state $\sigma$. Integrating Eq.~\eqref{subeq:dtd-fp-out} over time shows that the normalization condition \eqref{eq:dtd-norm} stems from the fact that $p_{\infty}^{(\text{out})}(\sigma) = 1$, i.e., the motor eventually leaves the current binding site.
 
We determine the dwell-time distributions from the probability fluxes as
\begin{equation}  
    p(\Omega,t|\sigma) = \frac{j(\Omega,t|\sigma)}{p(\Omega|\sigma)} \ , \quad \Omega = \{{\rm f}, {\rm b}\} \ .
\end{equation}
The existence of four dwell-time distributions---corresponding to forward and backward transitions from the two possible initial chemical states $\sigma$---has a striking implication: 
At a coarse-grained level, the system follows a \textit{second-order Markov process}, rather than a first-order Markov process as assumed in purely chemical-kinetic descriptions (schematically depicted in Fig.~\ref{fig:kinetic-approach}), because the initial chemical state depends on the immediately preceding binding site from which the motor arrived. The future evolution depends on the current binding site and the immediately previous one.

Importantly, two of the four probability fluxes in Eq.~\eqref{eq:dwell-time-dis} are related by microscopic irreversibility~\cite{crooks_entropy_1999}:
\begin{equation}\label{eq:ratio}
    \frac{j({\rm f},t|0)}{j({\rm b},t|1)} = e^{2\beta\Delta \mathcal{F}} \, .
\end{equation}
This implies that the dwell-time distributions $p({\rm f},t|0)$ and $p({\rm b},t|1)$ are identical. Physically, the ensemble of trajectories contributing to $j({\rm f},t|0)$ is the time-reversal of the ensemble contributing to $j({\rm b},t|1)$, so that their probability ratio is governed by the total entropy production in the environment. In this case, it equals twice the free energy per completed step, since two successive head transitions are involved in a full motor step. By contrast, an analogous relation does not hold for the probability fluxes $j({\rm f},t|1)$ and $j({\rm b},t|0)$, as the corresponding trajectory ensembles are not related under time reversal. Consequently, the distributions $p({\rm f},t|1)$ and $p({\rm b},t|0)$ are generally different.

\begin{figure}
\centering
\includegraphics[width=0.47\textwidth]{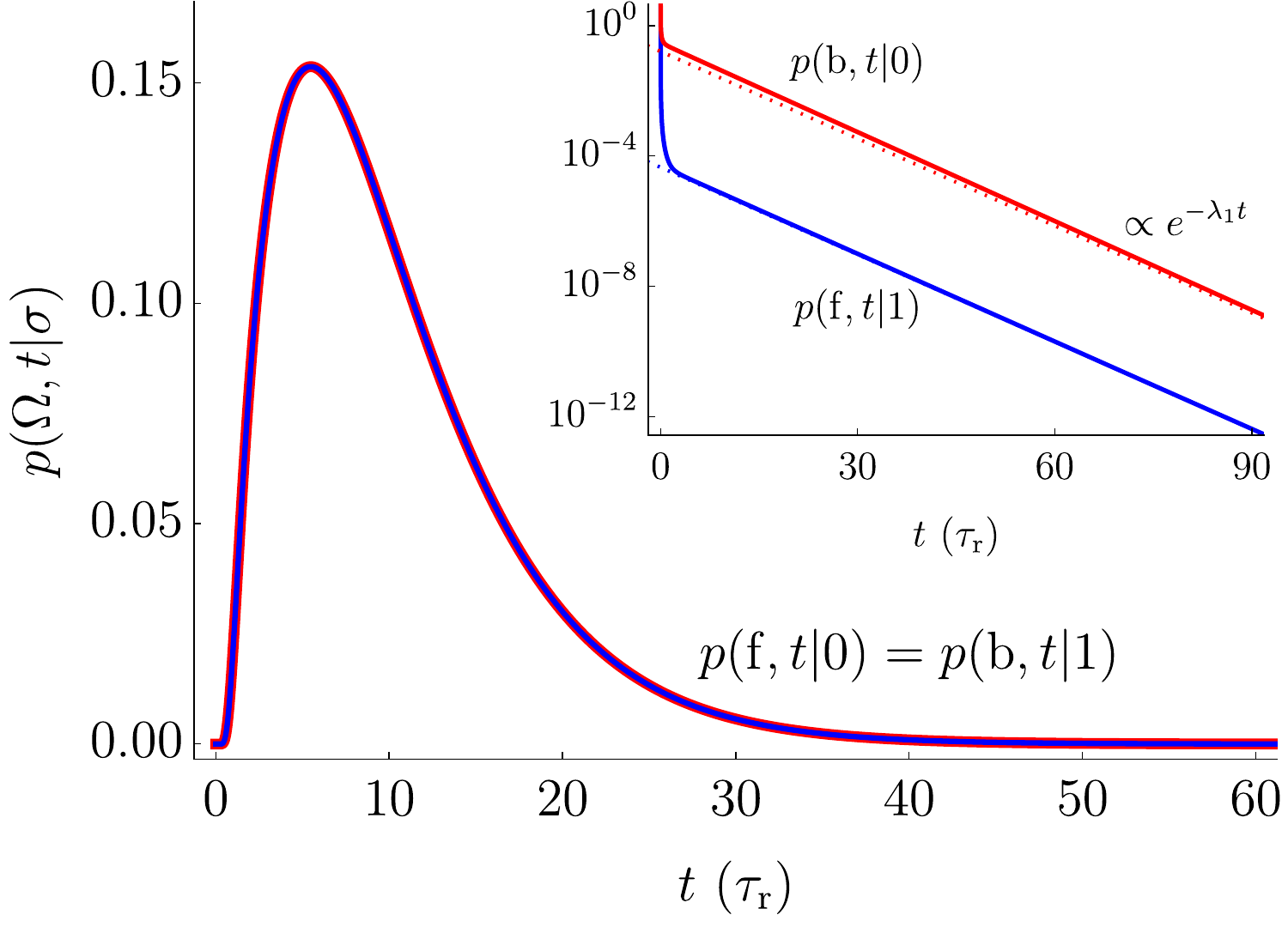}
\caption{Dwell-time distributions for forward and backward motor stepping, conditioned on the initial chemical state, $0$ (red) or $1$ (blue), at the current binding site. Throughout, we consider bare switching rate $k_0 = \tR^{-1}$, chemical free energy $\Delta \mu = 10 \, \kT$, step length $a = \sigma_{\rm eq}$, and cargo force $f = f^*/2$, with harmonic confinement $V(r) = \tfrac{1}{2} \kappa r^2$. Main panel: non-monotonic distributions $p({\rm f},t|0) = p({\rm b},t|1)$ in linear scale. Inset: sharply peaked distributions $p({\rm b},t|0)$ and $p({\rm f},t|1)$ on log scale. Dashed straight lines: long-time exponential limit characterized by the dominant eigenvalue $\lambda_1$.}
\label{fig:waiting-time-dis}
\end{figure}

Figure~\ref{fig:waiting-time-dis} shows the four dwell-time distributions for a representative choice of model parameters and a harmonic interaction $V(r) = \tfrac{1}{2}\kappa r^2$. The two forward-step distributions (and the two backward-step distributions) qualitatively differ based on the initial state $\sigma$. The distributions $p({\rm f},t|1)$ and $p({\rm b},t|0)$ describe transitions starting at the target mechanical state (a binding site) where the switch can occur immediately, therefore they are sharply peaked at $t = 0$. In contrast, the distributions $p({\rm b},t|1) = p({\rm f},t|0)$ are non-monotonic, with zero support at $t = 0$ and instead maximized at intermediate time $t$, because the motor head must first move to the target location before the switch can occur. 

The behavior of these distributions can be inferred from the spectral decomposition of the Fokker--Planck operator. In particular, the solution to Eq.~\eqref{eq:motor-kinetics-waiting-time} can be expressed as an eigenmode expansion~\cite{risken_fokker_1989},
\begin{equation}\label{eq:eigen}
    P^{(\text{in})}_t(\nu,r|\sigma) = \sum_{{\ell}=1}^{\infty} c^{(\ell)}(\sigma) \, \phi^{(\ell)}(\nu,r)\, e^{-\lambda_{\ell} t} \ ,
\end{equation}
for orthonormal eigenfunctions $\phi^{(\ell)}(\nu,r)$ and corresponding eigenvalues $\lambda_{\ell}$ of the Fokker–Planck operator. The coefficients $c^{(\ell)}(\sigma)$ are determined by projecting the initial condition onto the eigenbasis. For the Fokker–Planck operator~\eqref{eq:FP-operator} with a confining potential $V(r)$, the spectrum is typically discrete and nondegenerate, so that the eigenvalues can be ordered as $\lambda_1 < \lambda_2 < \cdots$. At sufficiently long times, the relaxation dynamics are dominated by the leading eigenmode ($\ell=1$), yielding
\begin{equation}
    P^{(\text{in})}_t(\nu,r|\sigma) \sim c^{(1)}(\sigma) \, \phi^{(1)}(\nu,r)\, e^{-\lambda_1 t} \ , \quad t \gg \lambda_1^{-1}.
\end{equation}
Since $P^{(\text{in})}_t(\nu,r|\sigma)$ is directly related to the dwell-time distributions through Eq.~\eqref{eq:dwell-time-dis}, it follows that all dwell-time distributions decay exponentially for sufficiently long times, with relaxation rate given by the smallest eigenvalue $\lambda_1$ (Fig.~\ref{fig:waiting-time-dis}). 

As discussed in Sec.~\ref{sec:dwell-time-dis}, purely exponential dwell-time distributions arise in chemical kinetic models described by continuous-time Markov jump processes, where the transition rate out of a state is independent of the time already spent in that state. Although our model dynamics are inherently non-Markovian, at sufficiently long times the system has explored a large portion of the phase space, and information about the initial chemical state is effectively lost, recovering the chemical-kinetic picture with exponential dwell times. The timescale separation from Eq.~\eqref{eq:timescale-separation} explicitly assumes sufficient time to travel to the target location and switch mobilities, thus supporting the consistency of our results.

\section{Discussion}
We have studied the dynamics of the two coordinated heads of dimeric transport motors, introducing a model of coupled overdamped Brownian motion with mobilities alternating in a position-dependent manner. Rather than relying on an explicit mechanical driving force, this stepping mechanism is that of a pure information engine, where coordinated mobility switching harnesses thermal fluctuations to generate directed motion. 

At the ensemble level, the dynamics are described by two coupled Fokker--Planck equations (corresponding to the two mobility states) with source and sink terms modeling stochastic switching of mobility. The transition rates between mobility states are chosen such that (i) the two heads are symmetric, and (ii) they satisfy local detailed balance, where chemical free energy is consumed during forward motion. A change of variables decouples the overall translational motion of the motor from its internal steady-state dynamics, which are uniquely characterized by the distribution of the relative displacement of the two motor heads, conditional on $\nu$. In particular, all relevant mechanical observables and thermodynamic flows in the steady state depend solely on the relative-displacement fluxes.  

Switching is localized in many biological contexts (e.g., kinesin binding to microtubules and myosin binding to actin); in our model such localized switching allows for substantial analytical progress and physical insight. At any given time one motor head (e.g., $x_1$) is fixed, while the other is free to move, starting from $x_0 = x_1-a$ and switching mobility only upon reaching either $x_0 = x_1 +a$ (a forward switch) or $x_0 = x_1-a$ (a backward switch). The mean displacement $\overline{\Delta r}$ of a motor head between switches therefore equals the step length $2a$ weighted by the imbalance between the respective forward and backward switch probabilities, $p({\rm f})$ and $p({\rm b})$, whereas the mean duration $\overline{\Delta t}$ of a head step is the inverse of the sum of the steady-state rates for forward and backward switches. For such localized switching, the stall force takes a particularly simple form, given by the chemical free energy per head step, $f_{\rm s} = \Delta \mu/(2a)$, in agreement with experimental observations for myosin-V~\cite{howard_mechanics_2002,mehta_myosin-v_1999}. 

Thermodynamically, our model behaves as a pure information engine, in which on average no net energy flows between the chemical and mechanical degrees of freedom. Focusing on the dynamics of a single motor head (without loss of generality, $x_0$), the chemical degree of freedom conditioned on fixed $x_1$ acts effectively as a Maxwell demon, switching only when $x_0$ reaches a binding site. From the point of view of $x_0$, once it becomes the moving head again after two switches, $x_1$ has effectively changed instantaneously (i.e., without any accompanying motion of $x_0$) in such a way that on average $x_1$ has done no work on $x_0$. This mechanism is analogous to that of non-autonomous stochastic engines consisting of a Brownian particle driven by an externally controllable optical trap~\cite{saha_maximizing_2021,patron_castro_harnessing_2026}. Since $x_0$ receives no average work (and exactly no work for localized switching), its motion relies solely on thermal fluctuations to reach binding sites and trigger subsequent mobility switching, resulting in directed motion accompanied by net heat uptake from the thermal environment.
Our results thus support the view that information transduction, rather than internal energy transmission, may underlie the remarkable performance of dimeric molecular motors.

Under localized switching, our model also admits a discrete-state description. This model presents two significant differences from (closely related) traditional chemical kinetic models. First, local detailed balance is not satisfied at the coarse-grained level. This violation arises from entropic production associated with the coarse-grained (mechanical) degrees of freedom, and its scale depends on the interplay between multiple intrinsic timescales in the dynamics: the diffusive times for mechanical motion between adjacent binding sites and the mean residence times that the motor head spends around a binding site before switching. Purely discrete-state models typically assume a clear separation of timescales, in which chemical switches are much slower than mechanical relaxation, allowing the system to effectively explore phase space between successive chemical events. This assumption implies that the dwell-time distributions are exponential at all times, with the same relaxation rates. In the opposite limit (for fast chemical dynamics), the motor attains biased motion even when the kinetic distinction breaks down and the probabilities for forward and backward motion become equal. 

Second, the discrete-state dynamics are described by a second-order Markov process. In this case, the future evolution of the system depends not only on its current state (as in a standard Markovian description) but also on the immediately preceding state. This feature is also present in chemical-network models for kinesin~\cite{valleriani_dwell_2008}.  As a consequence, four distinct dwell-time distributions arise, corresponding to forward and backward motion conditioned on the two possible initial chemical states.

The model’s non-Markovian character provides valuable insights and experimentally testable predictions. Experimental analysis based on purely discrete-state models typically reports only two dwell-time distributions, associated with forward and backward steps, effectively coarse-graining over the internal dynamics. However, using the same experimental trajectories, the four dwell-time distributions predicted here can in principle be extracted by measuring the residence time in a given metastable state (binding site) conditioned on the previous metastable state from which the motor arrived. This perspective opens a new window for inferring microscopic energetics of individual motor heads from coarse-grained stepping statistics. For example, the linker potential between the two heads influences the shape of the dwell-time distributions, suggesting that these observables could provide indirect information about the underlying mechanics.

Here we treated the cargo in a simplified manner of a constant opposing force. A natural extension would be to incorporate explicit cargo dynamics as an additional degree of freedom coupled to the motor, and to investigate how this affects both mechanical observables and thermodynamical flows. Previous theoretical studies have shown that approximating the effect of the cargo by a constant force can lead to an artifactually larger variance in the motor step number compared to models that explicitly account for cargo diffusion~\cite{brown_pulling_2019}. Moreover, our motor model assumes high processivity, such that complete detachment is not allowed, even under strong opposing forces. Exploring the role of track-head interactions at moderate-to-large forces would therefore be of particular interest, especially given that experimental motor velocities are typically calculated only from trajectories in which the motor remains attached to the cargo, effectively filtering out contributions from the large-force regime.

\begin{acknowledgments}
This work was supported by a Natural Sciences and Engineering Research Council (NSERC) Discovery Grant RGPIN-2020-04950, an NSERC Alliance International Collaboration Grant ALLRP-2023-585940, and a Tier-II Canada Research Chair CRC-2020-00098 (all D.A.S.). 
\end{acknowledgments}

\section*{Data Availability}
The codes that support the findings of
this Letter are openly available~\cite{apatron_github_2026}.

\bibliography{SMP_bibliography}
\end{document}